\documentclass[english,12pt,a4paper]{article}
\usepackage{authblk}
\usepackage[text={17.0cm,23.7cm}]{geometry}
\usepackage{babel}
\usepackage{hyperref}
\usepackage{cite}
\usepackage[dvipsnames]{xcolor}
\usepackage{graphicx}
\usepackage{amsmath}
\usepackage{amssymb}
\usepackage{xcolor}
\usepackage{yfonts}
\usepackage{dsfont}
\usepackage{booktabs}
\usepackage{braket}
\usepackage{ulem}
\usepackage{placeins}
\usepackage{afterpage}
\usepackage{mathrsfs}
\usepackage[mathscr]{euscript}

\newcommand{\CA}{C_\text{A}}
\newcommand{\CE}{C_\text{E}}

\newcommand{\mTi}{m_{\text{T}_i}}

\newcommand{\GammaA}{\Gamma_{\text{A}}}

\begin{document}
	\title{\vspace{-2cm}
		{\normalsize
			\flushright TUM-HEP 1295/20\\}
		\vspace{0.6cm}
		\textbf{Conservative constraints on the effective theory of dark matter-nucleon interactions from IceCube: the impact of operator interference}\\[8mm]}

	\author[a,b]{Anja Brenner}
	\author[b]{Alejandro Ibarra}
	\author[b]{Andreas Rappelt}
	\affil[a]{\normalsize\textit{Max-Planck-Institut f\"ur Physik (Werner-Heisenberg-Institut),\protect\\ F\"ohringer Ring 6, 80805 M\"unchen, Germany}}
	\affil[b]{\normalsize\textit{Physik-Department, Technische Universit\"at M\"unchen, \protect\\James-Franck-Stra\ss{}e, 85748 Garching, Germany}}

	\date{}

\maketitle
\begin{abstract}
We present a method to derive conservative upper limits on the coupling constants of the effective theory of dark matter-nucleon interactions, taking into account the interference among operators. The method can be applied in any basis, and can be easily particularized to any UV complete model. To illustrate our method, we use the IceCube constraints on an exotic neutrino flux from dark matter annihilations in the Sun to derive conservative upper limits on the dark matter-nucleon coupling constants of the effective theory, as well as to derive conservative upper limits on the dark matter-proton and dark matter-neutron scattering cross-sections.
\end{abstract}

\section{Introduction}
\label{sec:intro}

Despite the enormous number of observations hinting towards the existence of dark matter in our Universe, the nature of this substance remains a mystery to this day. A plausible hypothesis is that the dark matter is constituted by cosmologically long-lived neutral particles, not contained in the Standard Model of Particle Physics, that interact with the ordinary matter (see {\it e.g.} \cite{Bertone:2004pz,Bergstrom:2000pn,Bertone:1900zza}). Under this assumption, one may expect annihilations of dark matter particles in overdense regions of dark matter, generating a flux of Standard Model particles that may be large enough to be disentangled from the astrophysical backgrounds. This search strategy is commonly known as indirect detection.

Optimal targets for indirect dark matter detection should have large dark matter concentrations, should be close to the detector, and should generate small background fluxes. Prime candidates for indirect detection are the Milky Way center and dwarf spheroidal galaxies, where the dark matter concentration is large. On the other hand, it was pointed out long time back that the Sun could also be a promising target for indirect detection~\cite{Silk:1985ax,Srednicki:1986vj,Griest:1986yu}. Dark matter particles could scatter off the matter of the solar interior, lose energy, and therefore become gravitationally trapped, generating after subsequent scatterings an overdense region of dark matter particles where annihilations can occur. The total annihilation rate in the Sun is clearly smaller than in the Galactic Center region. Nevertheless, the proximity of the Sun to the terrestrial detectors may lead to detectable fluxes.

The electromagnetically and/or strongly interacting particles produced in the annihilation are stopped by the solar matter. However, high-energy neutrinos can escape the Sun and reach the Earth. A search for high-energy neutrinos in the direction of the Sun has been conducted by IceCube~\cite{Aartsen:2012kia,Aartsen:2016zhm}, Super-Kamiokande~\cite{Tanaka:2011uf}, ANTARES~\cite{Adrian-Martinez:2016gti} and Baikal~\cite{Avrorin:2014swy}. The non-observation of a significant excess of neutrinos with respect to the atmospheric neutrino background allows to set an upper limit on the size of an exotic flux component. For a given annihilation channel, and under the plausible assumption that the rate of capture is equal to the rate of annihilation, this upper limit is translated into limits on the spin-independent and spin-dependent interaction cross-section with the proton.

It is common in these works to impose that the dark matter interaction with the nucleons is isoscalar, {\it i.e.} it is identical for protons and for neutrons. Clearly, this assumption is very restrictive, and in fact in many scenarios this assumption does not hold (see {\it e.g.} \cite{Feng:2011vu,Gao:2011bq,Gao:2011ka,Frandsen:2011cg,Hamaguchi:2014pja,Belanger:2013tla,Drozd:2015gda,Martin-Lozano:2015vva}). Therefore, great care should be taken when confronting the predictions of these models with the upper limits reported by the experiments, as points which lie above the ``upper limit" on the cross-section may in fact be allowed by the experimental upper limit on the rate, as the result of an interference between the isoscalar and the isovector interactions. More appropriately, a limit on the cross-section with protons should be calculated for the predicted contributions from the isoscalar and isovector interactions in that given model, or more conservatively, the theoretical expectations should be compared to the largest value of the cross-section which is allowed (for that dark matter mass) among all possible interferences between the isoscalar and isovector contributions. 

Further, it has been emphasized in recent years that dark matter particles could interact with the nucleons through a larger set of operators, with nuclear response functions which can be different to those from the spin-independent and spin-dependent interactions~\cite{Fan:2010gt, Fitzpatrick:2012ix}. Limits on the size of the coupling strengths of the effective theory have been derived from the IceCube data in \cite{Catena:2016kro}, assuming that only one operator contributes to the capture in the Sun, and assuming that the interaction is either isoscalar or isovector. Also in this case, for a given model the interference between the isoscalar or isovector interaction can lead to rates in neutrino telescopes smaller than the upper limits derived under the assumption that one of them vanishes. Moreover, different forms of the interaction can interfere with one another, possibly leading to even smaller detection rates. In the effective theory of dark matter-nucleon interactions, the necessity of deriving conservative limits on the interaction cross-section, including the possibility of interferences among interactions, is therefore strengthened. 

In this paper we will develop a methodology to determine conservative upper limits on the coupling strengths of the effective theory of dark matter-nucleon interactions, including the effect of interferences. Here we will apply our method to derive limits on the coupling strengths from the non-observation of an exotic neutrino flux in the direction of the Sun using the three-year IceCube data, although the method is general and can be also applied to direct detection experiments, namely experiments searching for dark matter-induced nuclear recoils.

The paper is organized as follows. In section \ref{sec:preliminaries} we describe the formalism to determine the neutrino flux from dark matter annihilations inside the Sun and its relation to the capture in the effective theory of dark matter-nucleon interactions. In Section \ref{sec:method} we introduce our method to calculate the absolute maximum of a given coupling strength from the experimental upper limit on the capture rate, assuming equilibrium between dark matter capture and annihilation. In Section \ref{sec:results} we use the upper limits from the IceCube detector on the capture rate in the Sun to determine conservative upper limits on the coupling strengths of the effective theory of dark matter-nucleon interactions (in the isoscalar-isovector basis and in the neutron-proton basis), as well as on the  spin-independent (SI) and spin-dependent (SD) dark matter-nucleon interaction cross-sections. Finally, in Section \ref{sec:conclusions} we present our conclusions. We also include Appendices providing more details about the complementarity of the different elements in the Sun in probing the parameter space of the effective field theory of dark matter-nucleon interactions, and 	generalizing the methodology of Section \ref{sec:method} for the case where equilibrium between capture and annihilation is not attained.

\section{High energy neutrinos from dark matter annihilations inside the Sun}
\label{sec:preliminaries}

\indent The Sun is being constantly bombarded by dark matter particles. Most dark matter particles traverse the Sun, however, a dark matter particle can sporadically  scatter off a nucleus in the solar interior. If the speed of the dark matter particle after the scattering is smaller than the escape velocity at the position of the scattering, the dark matter particle would become gravitationally bound to the Sun and eventually sink to the center after subsequent scatterings~\cite{Press:1985ug}.
The number of dark matter particles trapped in the solar core, on the other hand, decreases due to annihilation or evaporation. The time evolution of the number of dark matter particles trapped inside the Sun,  $N(t)$, is governed by the following differential equation:
\begin{align}
\frac{dN}{dt}=C- \CE\, N- \CA \,N^2.
\label{eq:diff_eq_number}
\end{align}
Here, $C$ is the capture rate, $\CE$ is the evaporation rate and $\CA$ is the annihilation coefficient. Dark matter evaporation is relevant for dark matter lighter than 1 - 4 GeV \cite{busoni2017evaporation}. In this work we will be interested in dark matter particles heavier than 20 GeV, therefore we will neglect the effects of evaporation. In this case, the solution to Eq.~(\ref{eq:diff_eq_number}) reads
\begin{align}
N(t)\,=\,\sqrt{\frac{C}{\CA}}\,\tanh\left(\frac{t}{\tau}\right)\;,
\label{eq:DM_number}
\end{align}
where $\tau  \,\equiv \,1/\sqrt{C\,\CA}$ denotes the equilibration time.

Finally, the differential neutrino flux from the annihilation of dark matter particles captured in the interior of the Sun reads
\begin{align}
\frac{d\Phi_\nu}{dE_\nu}=\frac{\Gamma_{A}}{4\pi d^2}\sum_i {\rm BF}_i\frac{dN^i_\nu}{dE_\nu}\;.
\label{eq:nuflux}
\end{align}
Here, $d=1.5\times 10^{11}\,{\rm m}$ is the distance between the Sun and the Earth,  $dN^i_\nu/dE_\nu$ is the energy spectrum of neutrinos produced in the annihilation channel $i$ with branching ratio ${\rm BF}_i$, and $\GammaA$ is the total annihilation rate at the present time, given by
\begin{align}
\GammaA\,=\,\frac{1}{2}\,\CA N(t)^2\,=\,\frac{1}{2}\,C\,\tanh^2\left(\frac{t_\odot}{\tau}\right)\;,
\label{eq:relation_ann_cap_rate}
\end{align}
where $t_\odot=4.6\times 10^9$ yr is the age of the Sun.

Both the capture rate and the annihilation constant are related to the fundamental parameters of the dark sector through the dark matter mass $m_\chi$, the total annihilation cross-section $\sigma v$, and the  differential scattering cross-section of a dark matter particle with the nucleus $i$, $d \sigma_i/d E_R$, with $E_R$ the recoil energy of the nucleus. For the Sun, the annihilation constant can be approximated by \cite{Jungman:1995df}
\begin{align}
\CA\simeq 1.63\times 10^{-52}\,{\rm s}^{-1}
\left(\frac{(\sigma v )}{3\times 10^{-26}\,{\rm cm}^3\,{\rm s}^{-1}}\right)\left(\frac{m_\chi}{\rm TeV}\right)^{3/2}\;.
\label{eq:CA_approx}
\end{align}

The capture rate, on the other hand, reads \cite{Gould:1987ir}
\begin{align}
C=& \sum_i \int_0^{R_\odot} \, d r \, 4\pi\,r^2\,\eta_i(r)\,\frac{\rho_\text{loc}}{m_\chi}\,\int_{v \leq v_{\text{max},i}^{\text{(Sun)}}(r)} d^3 v \, \frac{ f (\vec{v})}{v}\,w^2(r) \nonumber\\&			 		~\times\int_{m_\chi v^2 /2}^{2 \mu_{i}^2 w^2(r)/\mTi} dE_R \, \frac{d \sigma_i}{dE_R}(w(r), E_R) \;.
\end{align}
Here, $\eta_i (r)$ is the number density of the nucleon species $i$ at distance $r$, for which we adopt the Standard Solar Model AGSS09ph \cite{serenelli2009new}. Further, $v$ denotes the dark matter velocity asymptotically far away from the gravitational source, and $w^2 (r)\,=\,v^2\,+\,v_{\mathrm{esc}}^2 (r)$
is the dark matter velocity at the distance $r$, where $v_{\mathrm{esc}}(r)$ is the escape velocity at that distance.
Finally,  $\rho_{loc}$ and $f(\vec{v})$ are the local dark matter  density and velocity distribution, for which we adopt the values of the Standard Halo Model, {\it i.e.} 0.3 GeV/cm$^3$ for the former \cite{read2014local, cerdeno2010direct}, and a Maxwell-Boltzmann distribution for the latter (with velocity dispersion 220 km/s \cite{Xue:2008se, McMillan:2009yr, Bovy:2009dr} and escape velocity 544 km/s \cite{Smith:2006ym, Piffl:2013mla} from the Milky Way).

Since dark matter particles are expected to move non-relativistically in the Solar System, we will consider for our analysis the most general form of the dark matter-nucleon scattering cross-section compatible with the Galilean symmetry. As shown in \cite{Fan:2010gt, Fitzpatrick:2012ix}, the Hamilton operator can be expressed in terms of a set of basic operators invariant under Galilean transformations and Hermitian conjugation: the momentum transfer $\widehat{\bf q}$, the transverse velocity $\widehat{\bf{v}}^\perp\equiv \, \widehat{\bf{w}}+\bf q/2\mu_{\mathscr N}$ (with $\mu_{\mathscr N}$ and $\widehat{\bf{w}}$ the dark matter-nucleon reduced mass and relative velocity), the dark matter spin, $\widehat{\bf{S}}_\chi$, and the nucleon spin, $\widehat{\bf{S}}_{\mathscr N}$.
The most general Hamiltonian for dark matter-nucleus interactions reads
\begin{align}
	\widehat{\cal H}_{\chi {\cal N}} = \sum_a
	\sum_{i} {\cal C}_i^a\widehat{\mathcal{O}}_i^a\,,
\end{align}
where the index $a$ labels the nucleons in the target nucleus, $a=1, ..., A$, with $A$ the mass number, and $i$ labels the possible non-relativistic interaction types. Keeping terms at most linear in $\widehat{\bf v}^\perp$, there are 14 independent interaction types for particles with spin up to $1/2$, which are listed in Table \ref{tab:Operators}.~\footnote{The effective theory up to spin 1 was considered in \cite{Dent:2015zpa,Catena:2019hzw}, and for arbitrary spin in \cite{Gondolo:2020wge}. We will restrict ourselves to spin up to 1/2, although the methodology that will be presented in this paper can be straightforwardly applied to higher spins.}
Therefore $\widehat{\mathcal{O}}^a_i$ denotes the non-relativistic operator for interactions of type $i$ between the dark matter particle and the $a$-th nucleon. Further, ${\cal C}_i^a$ denotes the coupling strength of the operator $\widehat{\mathcal{O}}_i^a$. Since the nucleon is an isospin doublet, ${\cal C}_i^a$ can be expressed as a $2\times 2$ matrix. It is common to express 
\begin{align}
	{\cal C}_i^a=c_i^0 \mathds{1}^a_{2\times 2}+c_i^1 \mathds{\tau}_3^a\,,
\end{align}
where $\mathds{1}_{2\times 2}^a$ ($\mathds{\tau}_{3}^a$) is the identity (third Pauli matrix) in the $a$-th nucleon isospin space, and $c_i^0$ ($c_i^1$) is the associated isoscalar (isovector) coupling constant. Alternatively, one can cast
\begin{align}
	{\cal C}_i^a=c_i^p (\mathds{1}^a_{2\times 2}+\mathds{\tau}_3^a)
	+c_i^n (\mathds{1}^a_{2\times 2}-\mathds{\tau}_3^a)\;,
\end{align}
where
\begin{align}
	c_i^n&=\frac{1}{2}(c_i^0-c_i^1)\;,\nonumber \\
	c_i^p&=\frac{1}{2}(c_i^0+c_i^1)\; 
\label{eq:np_from_10}
\end{align}
are respectively the coupling constants to the neutron and the proton. Finally, from the interaction Hamiltonian, we calculate $d\sigma/dE_R$ using the methods described in~\cite{Fitzpatrick:2012ix,Catena:2015uha}. We use one-body density matrix elements (OBDMEs) computed in~\cite{Anand:2013yka} and implemented in the {\sffamily Mathematica} package {\sffamily DMFormFactor}. 

\begin{table}[t!]
	\centering
	\begin{tabular}{ll}
		\toprule
		\toprule
		$\widehat{\mathcal{O}}_1 = \mathds{1}_{\chi}\mathds{1}_{\mathscr N}$  & $\widehat{\mathcal{O}}_9 = i\widehat{\textbf{S}}_\chi\cdot\left(\widehat{\textbf{S}}_{\mathscr N}\times\frac{\widehat{\textbf{q}}}{m_{\mathscr N}}\right)$ \\
		$\widehat{\mathcal{O}}_3 = i\widehat{\textbf{S}}_{\mathscr N}\cdot\left(\frac{\widehat{\textbf{q}}}{m_{\mathscr N}}\times\widehat{\textbf{v}}^{\perp}\right)\mathds{1}_\chi$ & $\widehat{\mathcal{O}}_{10} = i\widehat{\textbf{S}}_{\mathscr N}\cdot\frac{\widehat{\textbf{q}}}{m_{\mathscr N}}\mathds{1}_\chi$ \\
		$\widehat{\mathcal{O}}_4 = \widehat{\textbf{S}}_{\chi}\cdot \widehat{\textbf{S}}_{\mathscr N}$ & $\widehat{\mathcal{O}}_{11} = i\widehat{\textbf{S}}_\chi\cdot\frac{\widehat{\textbf{q}}}{m_{\mathscr N}}\mathds{1}_{\mathscr N}$ \\
		$\widehat{\mathcal{O}}_5 = i\widehat{\textbf{{S}}}_\chi\cdot\left(\frac{\widehat{\textbf{{q}}}}{m_{\mathscr N}}\times\widehat{\textbf{{v}}}^{\perp}\right)\mathds{1}_{\mathscr N}$ & $\widehat{\mathcal{O}}_{12} = \widehat{\textbf{S}}_{\chi}\cdot \left(\widehat{\textbf{S}}_{\mathscr N} \times\widehat{\textbf{v}}^{\perp} \right)$ \\
		$\widehat{\mathcal{O}}_6 = \left(\widehat{\textbf{S}}_\chi\cdot\frac{\widehat{\textbf{q}}}{m_{\mathscr N}}\right) \left(\widehat{\textbf{S}}_{\mathscr N}\cdot\frac{\widehat{\textbf{{q}}}}{m_{\mathscr N}}\right)$& $\widehat{\mathcal{O}}_{13} =i \left(\widehat{\textbf{S}}_{\chi}\cdot \widehat{\textbf{v}}^{\perp}\right)\left(\widehat{\textbf{S}}_{\mathscr N}\cdot \frac{\widehat{\textbf{q}}}{m_{\mathscr N}}\right)$ \\
		$\widehat{\mathcal{O}}_7 = \widehat{\textbf{S}}_{\mathscr N}\cdot \widehat{\textbf{v}}^{\perp}\mathds{1}_\chi$ & $\widehat{\mathcal{O}}_{14} = i\left(\widehat{\textbf{S}}_{\chi}\cdot \frac{\widehat{\textbf{q}}}{m_{\mathscr N}}\right)\left(\widehat{\textbf{S}}_{\mathscr N}\cdot \widehat{\textbf{v}}^{\perp}\right)$\\
		$\widehat{\mathcal{O}}_8 = \widehat{\textbf{S}}_{\chi}\cdot \widehat{\textbf{v}}^{\perp}\mathds{1}_{\mathscr N}$  & $\widehat{\mathcal{O}}_{15} = -\left(\widehat{\textbf{S}}_{\chi}\cdot \frac{\widehat{\textbf{q}}}{m_{\mathscr N}}\right)\left[ \left(\widehat{\textbf{S}}_{\mathscr N}\times \widehat{\textbf{v}}^{\perp} \right) \cdot \frac{\widehat{\textbf{q}}}{m_{\mathscr N}}\right] $ \\
		\bottomrule
		\bottomrule
	\end{tabular}
	\caption{\small Operators defining the non-relativistic effective theory of dark matter interactions with the nucleon ${\mathscr N}=n,p$~\cite{Fan:2010gt, Anand:2013yka}.  In this list we have omitted $\mathcal{O}_2 =\widehat{\textbf{v}}^{\perp}\cdot\widehat{\textbf{v}}^{\perp}$, since it does not appear	at leading order in the effective theory of any relativistic interaction.}
	\label{tab:Operators}
\end{table}

One can collect the 28 coupling constants $c_i^0$ and $c_i^1$, $i=1,~...,~14$ in a single vector ${\bf c}$ with components $c_\alpha$, $\alpha=1,~...,~28$, 
which encodes the types and strength of interactions arising from a concrete dark matter model, upon matching to the effective theory of dark matter-nucleon interactions. The capture rate can then be written in the form~\cite{Catena:2016hoj,Catena:2018ywo}
\begin{align}
C\left({\bf c }\right) = {\bf c}^T \mathbb{C} {\bf c} \,,
\label{eq:C_quadratic}
\end{align}
where $\mathbb{C}$ is a $28\times 28$ real symmetric matrix, which depends on the dark matter mass, the solar model, and the local dark matter density and velocity distribution. This matrix encodes all the necessary information to analyze the implications of dark matter capture in neutrino telescopes in the effective theory of dark matter-nucleon interactions, as we will describe in the next section.

\section{Upper limit on a coupling constant from capture in the Sun including operator interference}
\label{sec:method}

The null search results from neutrino telescopes translate, for a given annihilation channel, into an upper limit on the total annihilation rate. Under the plausible assumption that the equilibration time is much shorter than the age of the Sun, the IceCube upper limits on the flux can be in turn translated into upper limits on the capture rate $C\leq C^{\rm u.l.}$, which depend on the dark matter mass and on the annihilation channel.

\begin{figure}[t!]
	\centering
	\includegraphics[width=.8\textwidth]{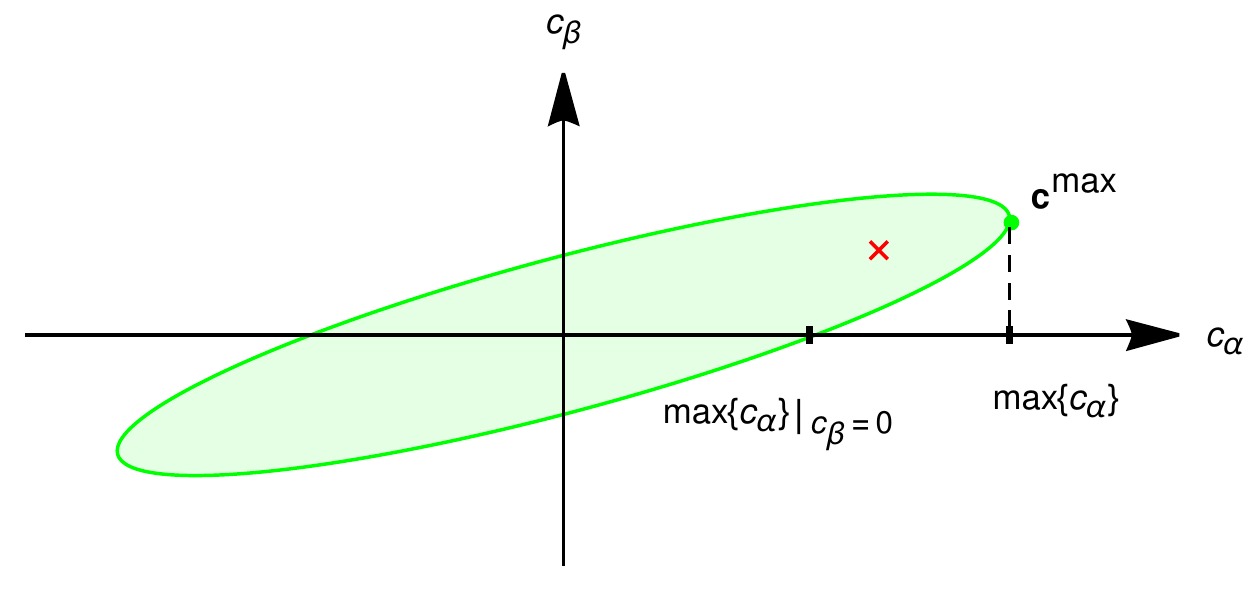}
	\caption{\small Region in the two-dimensional parameter space spanned by $c_\alpha$ and $c_\beta$ compatible with the condition $C\leq C^{\rm u.l.}$. The picture also shows the point compatible with the condition $C\leq C^{\rm u.l.}$  for which the coordinate $c_\alpha$ reaches the maximum value. This point is labeled as ${\bf c}^{\rm max}$, and has $c_\alpha$-coordinate equal to ${\rm max}\{c_\alpha\}$. The maximum value of the $c_\alpha$-coordinate compatible with the conditions $C\leq C^{\rm u.l.}$ and $c_\beta=0$ is labeled as ${\rm max}\{c_\alpha\}|_{c_\beta=0}$. The point labeled with a red cross is allowed by the condition $C\leq C^{\rm u.l.}$, and has $c_\alpha$-coordinate ${\rm max}\{c_\alpha\}|_{c_\beta=0}<c_\alpha<{\rm max}\{c_\alpha\}$.}
	\label{fig:ellipse}
\end{figure}

It is common in the literature to derive limits on the coupling constant $c_i^\tau$ from the requirement $C\leq C^{\rm u.l.}$ assuming that all other coupling constants vanish. However, this approach leads to too aggressive limits, and points violating this upper limits may in fact be allowed. This is illustrated in Fig.\ref{fig:ellipse}, which schematically shows the region in parameter space allowed by the requirement $C<C^{\rm u.l.}$, taking for simplicity a two-dimensional ${\bf c}$-space spanned by only two coupling constants $c_\alpha$ and $c_\beta$, although the argument can be extended for a 28-dimensional ${\bf c}$-space. The limit on the coupling constant $c_\alpha$ assuming $c_\beta=0$ is indicated as ${\rm max}\{c_\alpha\}|_{c_\beta=0}$. Clearly, there are points with $c_\alpha>{\rm max}\{c_\alpha\}|_{c_\beta=0}$ which satisfy the constraint $C<C^{\rm u.l.}$, {\it e.g.} the one indicated by a red cross. A rigorous upper limit on the coupling constant $c_\alpha$ (although conservative) is given by the point ${\rm max}\{c_\alpha\}$.

To calculate in full generality the maximum value of the $\alpha$-th component of the 28-dimensional vector ${\bf c}$ subject to the constraint $C < C^{\rm u.l.}$ we construct the Lagrangian
\begin{align}
L=c_\alpha-\lambda \Big[C({\bf c})-C^{\rm u.l.}\Big]\;,
\end{align}
with  $\lambda$ a Lagrange multiplier. The optimization conditions are
\begin{align}
\frac{\partial L}{\partial c_\beta}\Big|_{{\bf c}={\bf c}^{\rm max}}\,&=\,\delta_{\alpha\beta}\,-\,2\lambda\, \mathbb{C}_{\beta\zeta} c^{\rm max}_\zeta \,=\,0\;,
\label{eq:max_xi-equil}\\
\frac{\partial L}{\partial\lambda}\Big|_{{\bf c}={\bf c}^{\rm max}}\,&=
C({\bf c}^{\rm max})-C^{\rm u.l.}=0\;,
\label{eq:max_lambda-equil}
\end{align}
where ${\bf c}^{\rm max}$ is the point on the ellipse such that the coordinate $\alpha$ takes the maximal value: ${\rm max}\{c_\alpha\}=c^{\rm max}_\alpha$ (see Fig.~\ref{fig:ellipse}). From Eq.~(\ref{eq:max_xi-equil}) we obtain that the $\beta$-th coordinate of ${\bf c}^{\rm max}$ is:
\begin{align}
c_\beta^{\rm max}\,=\,\frac{1}{2\lambda} \left(\mathbb{C}^{-1}\right)_{\alpha\beta}\;.
\label{eq:c-max-equil}
\end{align}
Substituting in Eq.(\ref{eq:max_lambda-equil}) we obtain that the Lagrange multiplier is $\lambda=\frac{1}{2}\sqrt{C^{\rm u.l}/ \left(\mathbb{C}^{-1}\right)_{\alpha\alpha}}$. Finally, we find that the maximum value of the coupling constant $c_\alpha$ is:
\begin{align}
{\rm max}\{c_\alpha\}=c_\alpha^{\rm max}\,=\sqrt{\left(\mathbb{C}^{-1}\right)_{\alpha\alpha}\, C^{\rm u.l.} }\;.
\label{eq:max-c-equil}
\end{align}
From this expression, and using that the $\alpha$-th element of the vector ${\bf c}$ corresponds to the coupling constant $c_i^\tau$, for some $i$ and $\tau$, it follows ${\rm max}\{c_i^\tau\}$. 

Eq.~(\ref{eq:max-c-equil}) can also be applied when the maximization is done in a subspace of the full 28-dimensional space, for example, for a given Lorentz invariant model where only a subset of the 14 operators of Table \ref{tab:Operators} appears. Obviously, in this case the upper limit on the coupling constant becomes stronger. In particular, if only one operator contributes to the scattering, corresponding to the element $\alpha$ of the vector ${\bf c}$, 
\begin{align}
	{\rm max}\{c_\alpha\}|_{\substack{c_\beta=0}}=\sqrt{\left(\mathbb{C}_{\alpha\alpha}\right)^{-1}\, C^{\rm u.l.} }\;.
	\label{eq:c-max-diag}
\end{align}
Clearly
\begin{align}
	\frac{{\rm max}\{c_\alpha\}|_{\substack{c_\beta=0}}}{{\rm max}\{c_\alpha\}}=
		 \sqrt{\frac{(\mathbb{C}_{\alpha\alpha})^{-1}}{(\mathbb{C}^{-1})_{\alpha\alpha}}}\leq 1\;.
\end{align}
In some cases the factor $\sqrt{(\mathbb{C}_{\alpha\alpha})^{-1}/(\mathbb{C}^{-1})_{\alpha\alpha}}$ could be sizable, so the effect of the interference ought not to be neglected; we will show some examples in the next section.

Analogous expressions hold for a direct detection experiment, where the recoil/ionization rate is given by $R({\bf c})={\bf c}^T \mathbb{R} {\bf c}$, and for which the rate of recoils/ionizations is bounded from above by $R< R^{\rm u.l.}$. It is important to remark that in many direct detection experiments the target is mostly constituted by a single element, so there might be combinations of coupling strengths poorly bounded by the experiment ({\it i.e.} the corresponding ellipse in the $c_\alpha$ parameter space could be very elongated). The loss of sensitivity of a single target experiment to certain combinations of couplings can be avoided using several targets, with different responses to the isoscalar and isovector interactions, and which synergize to probe different parts of the parameter space. The Sun is a notable example of a multi-target experiment, where different elements (even if they have a small abundance) allow to probe regions of the parameter space for which the capture via scatterings with the most abundant elements $^1$H and $^4$He  is inefficient. For details, we refer to Appendix \ref{sec:complementarity}.

In some applications it may proof convenient to work in a different basis (for example, the neutron-proton basis instead of the isoscalar-isovector basis). Let us denote the alternative basis with a tilde.  From Eq.~(\ref{eq:max-c-equil}) it follows that
\begin{align}
	{\rm max}\{\tilde c_\alpha\}=\sqrt{(\widetilde{\mathbb{C}}^{-1})_{\alpha\alpha}\, C^{\rm u.l.} }\;,
	\label{eq:c-max-alt}
\end{align}
where $\widetilde{\mathbb{C}}$ can be calculated along similar lines as $\mathbb{C}$, but in the new basis. However, this matrix can be more easily calculated from the transformation between coupling constants
\begin{align}
	\tilde{\bf c}= \mathbb{T} {\bf c}\;,
\end{align}
and the condition that the capture rate is independent of the basis: 
\begin{align}
	C\left( {\bf c }\right) = {\bf c}^T \mathbb{C} {\bf c} \,=\tilde {\bf c}^T \widetilde{\mathbb{C}} \tilde{\bf c}\;.
	\label{eq:C_quadratic-basis}
\end{align}
Hence, one finds
\begin{align}
\widetilde{\mathbb{C}}=(\mathbb{T}^{\rm T})^{-1}\,\mathbb{C}\,\mathbb{T}^{-1}\;.
\label{eq:relation_C_Ctilde}
\end{align}

For instance, for a single operator ${\cal O}_i$, the two-dimensional vector of coupling constants in the isoscalar-isovector basis ${\bf c}=\{c_i^0,c_i^1\}$ is related to the corresponding vector in the neutron-proton basis $\tilde {\bf  c}=\{c_i^n,c_i^p\}$ by
\begin{align}
	\mathbb{T}=\frac{1}{2}\begin{pmatrix} 1 & -1 \\ 1 & 1 \end{pmatrix}\;
	\label{eq:T-matrix}
\end{align}
(see Eq.~(\ref{eq:np_from_10})). For the 28-dimensional vector of coupling constants, the corresponding $\mathbb{T}$-matrix would be a block-diagonal matrix with Eq.~(\ref{eq:T-matrix}) on the diagonal. Together with Eq.~(\ref{eq:relation_C_Ctilde}) and Eq.~(\ref{eq:c-max-alt}), the $\mathbb{T}$-matrix allows to calculate the upper limit on the coupling constants in the tilded basis, and in turn the limits on the dark matter-neutron and dark matter-proton scattering cross-sections.

\begin{figure}[t!]
	\centering
	\includegraphics[width=.6\textwidth]{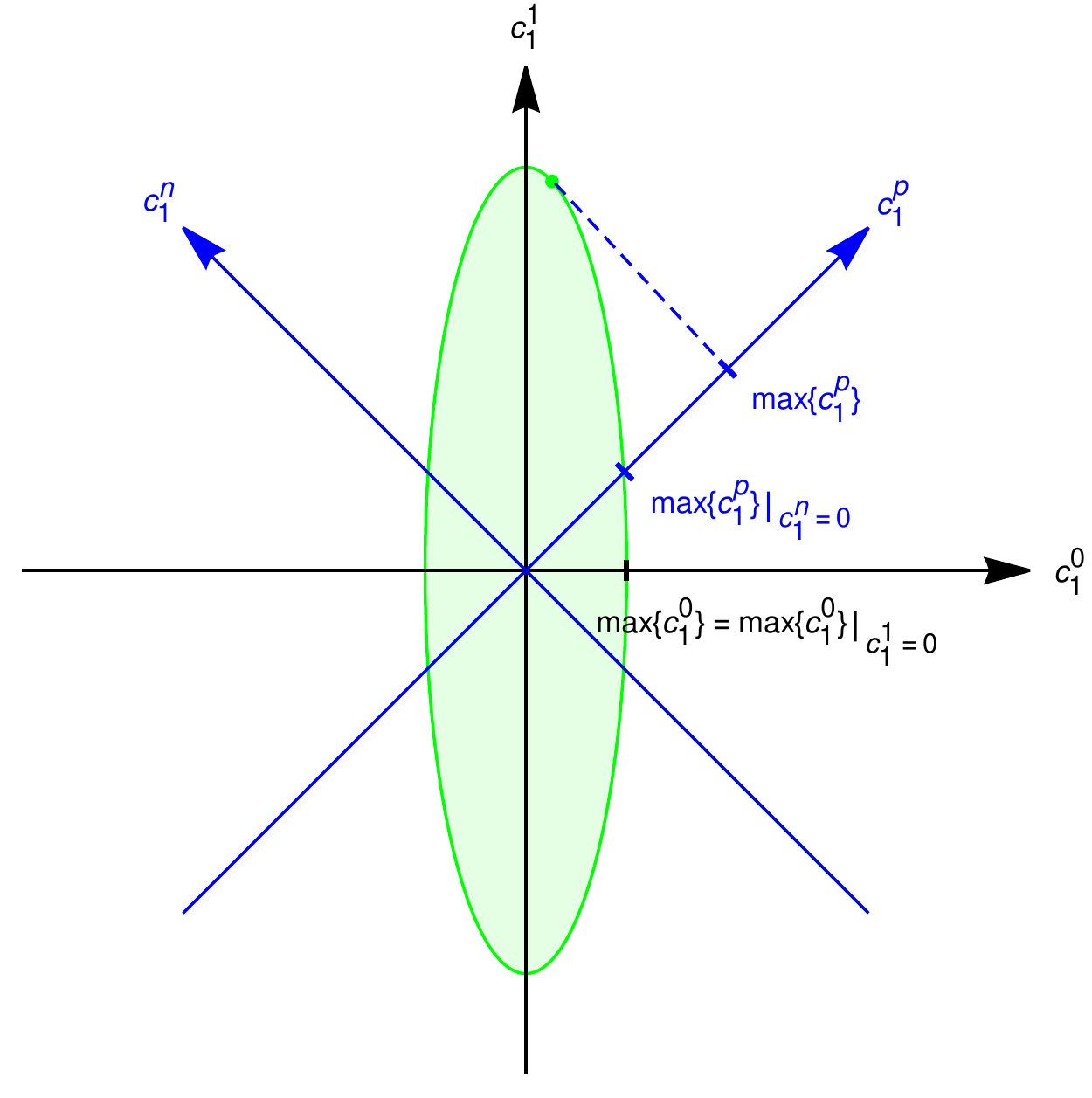}
	\caption{\small Schematic illustration of the impact of the interference between operators in different bases. The region in parameter space leading to a capture through the effective operator $\widehat {\cal O}_1$ compatible with the upper limit $C\leq C^{\rm u.l.}$ is depicted in green. In this particular example, the axes of the ellipse coincide with the axes of the isoscalar-isovector basis, $c_1^0, c_1^1$, and therefore the absolute maximum of the coupling strength $c_1^0$ coincides with the maximum calculated under the assumption $c_1^1=0$. Namely, ${\rm max}\{c_1^0\}={\rm max}\{c_1^0\}|_{c_1^1=0}$ and interference plays no role. In contrast, in the neutron-proton basis (shown in blue) the axes of the ellipse are misaligned with the corresponding axes in this basis, and correspondingly,  ${\rm max}\{c_1^p\}>{\rm max}\{c_1^p\}|_{c_1^n=0}$.}
	\label{fig:ellipse_bases}
\end{figure}

Let us note that the impact of the interferences in the derivation of the upper limits on the coupling constants can depend strongly on the chosen basis. This is illustrated in Fig. \ref{fig:ellipse_bases}, which shows the allowed region in the $\{c_1^0,c_1^1\}$-parameter space in the special case where the ellipse defined by  $C<C^{\rm u.l.}$ is aligned with the isoscalar and isovector axes. In this case, ${\rm max}\{c_1^0\}={\rm max}\{c_1^0\}|_{c_1^1=0}$,  ${\rm max}\{c_1^1\}={\rm max}\{c_1^1\}|_{c_1^0=0}$. However, in the neutron-proton basis, ${\rm max}\{c_1^n\}> {\rm max}\{c_1^n\}|_{c_1^p=0}$,  ${\rm max}\{c_1^p\}>{\rm max}\{c_1^p\}|_{c_1^n=0}$.

The method can be extended to the case where equilibration between capture and annihilation is not attained; this discussion is deferred to Appendix \ref{sec:no_equil}.

\section{Conservative constraints on the effective theory of dark matter nucleon interactions from IceCube}
\label{sec:results}

The IceCube Neutrino Observatory has undertaken a search for dark matter annihilations in the Sun during the austral winters between May 2011 and May 2014, amounting to a total of 532 days \cite{Aartsen:2016zhm}. No significant excess was found over the atmospheric background, leading to an upper limit on the neutrino flux from dark matter annihilations inside the Sun. Assuming equilibration between capture and annihilations, the IceCube collaboration has published 90\% C.L. upper limits on the capture rate for dark matter masses between 20 GeV and 10 TeV, for the final states $W^+W^-$, $b\bar b$ and $\tau^+\tau^-$ \cite{Aartsen:2016zhm}.~\footnote{The assumption of equilibration is well justified in view of the current IceCube limits when the dark matter annihilation cross-section takes the thermal value. In other scenarios, in contrast, equilibration may not hold and the limits derived in this section will be weakened. For a detailed discussion, see Appendix \ref{sec:no_equil}.}

\begin{figure}[htbp]
	\begin{center}
		\includegraphics[width=.45\textwidth]{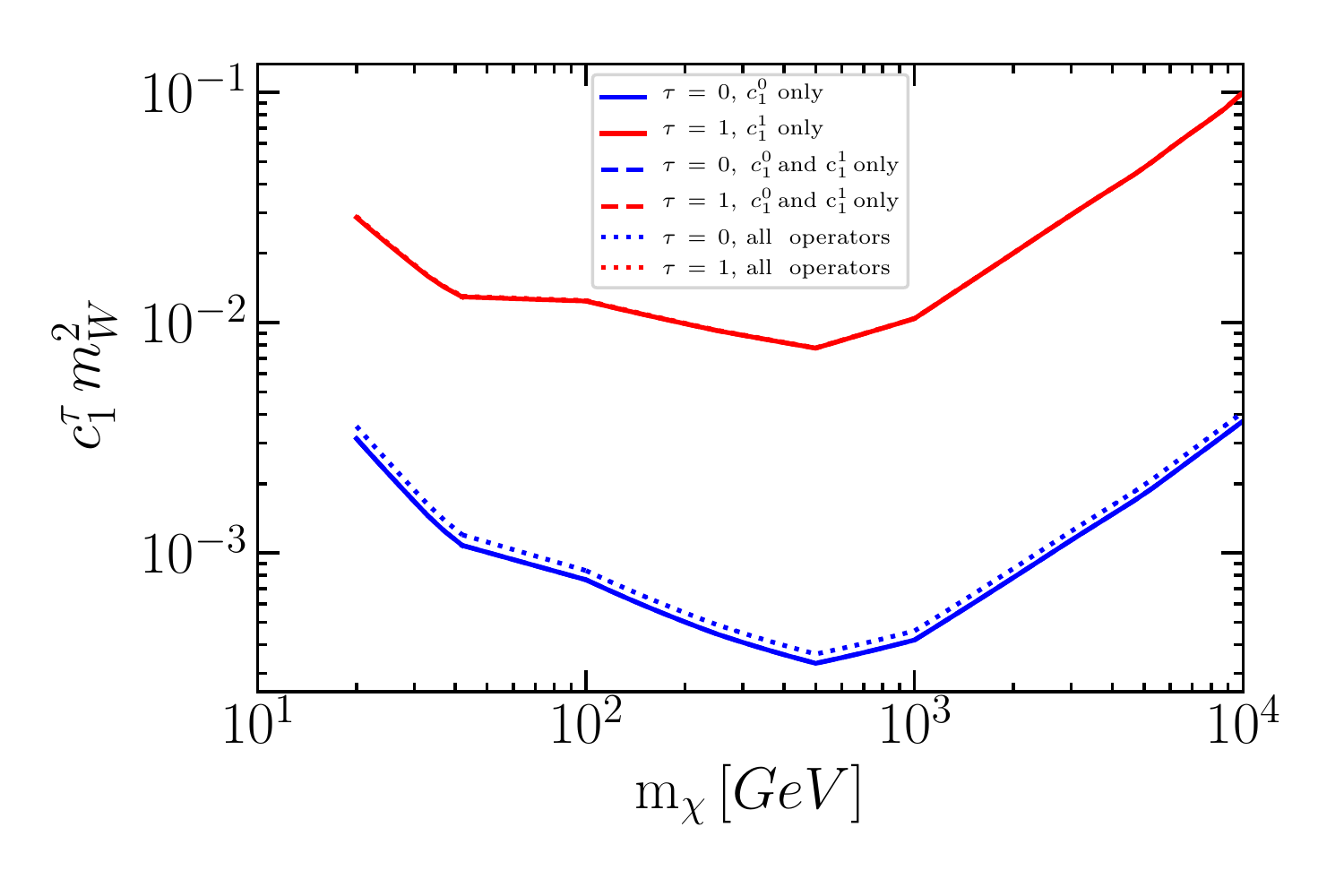}
		\includegraphics[width=.45\textwidth]{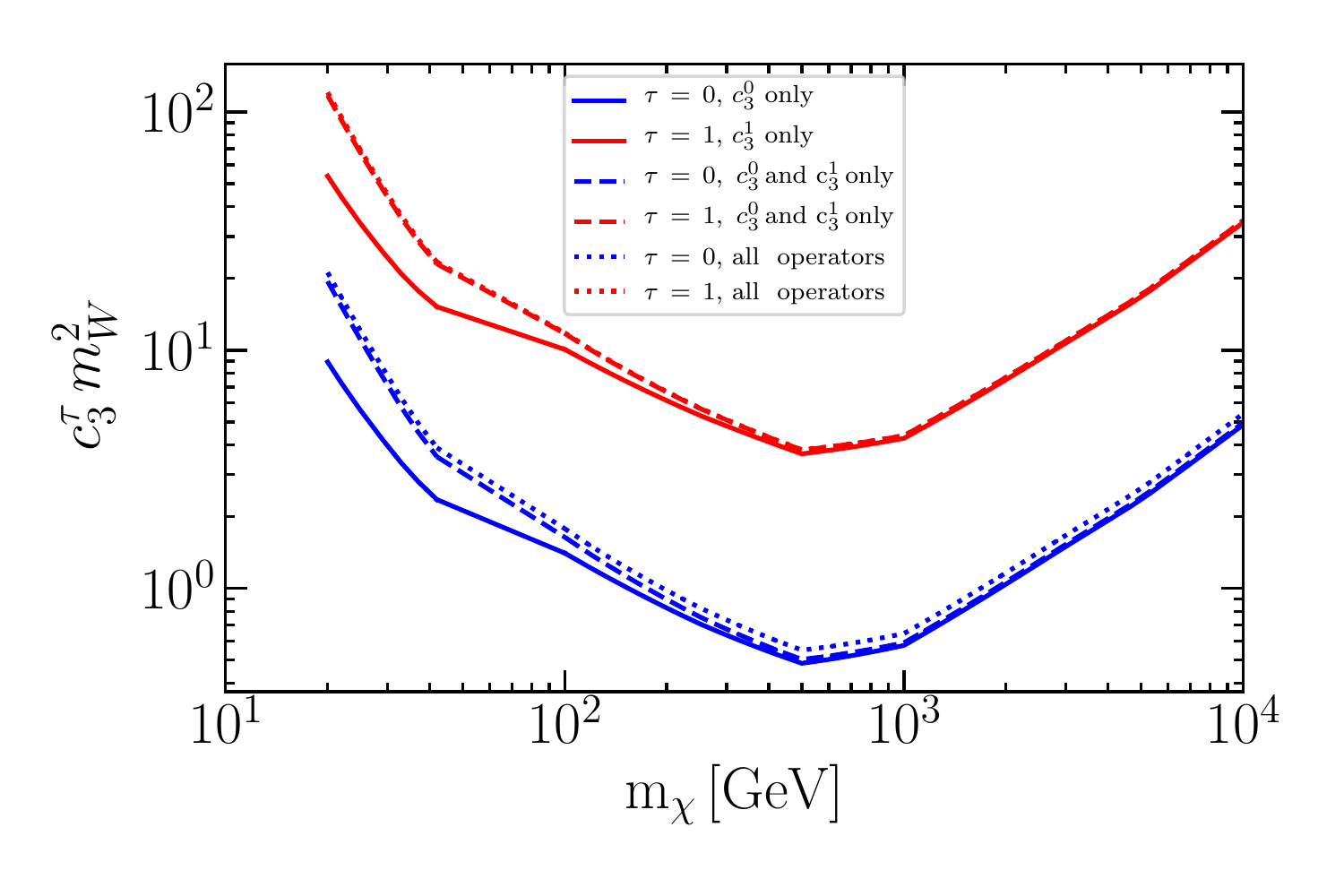}\\
		\includegraphics[width=.45\textwidth]{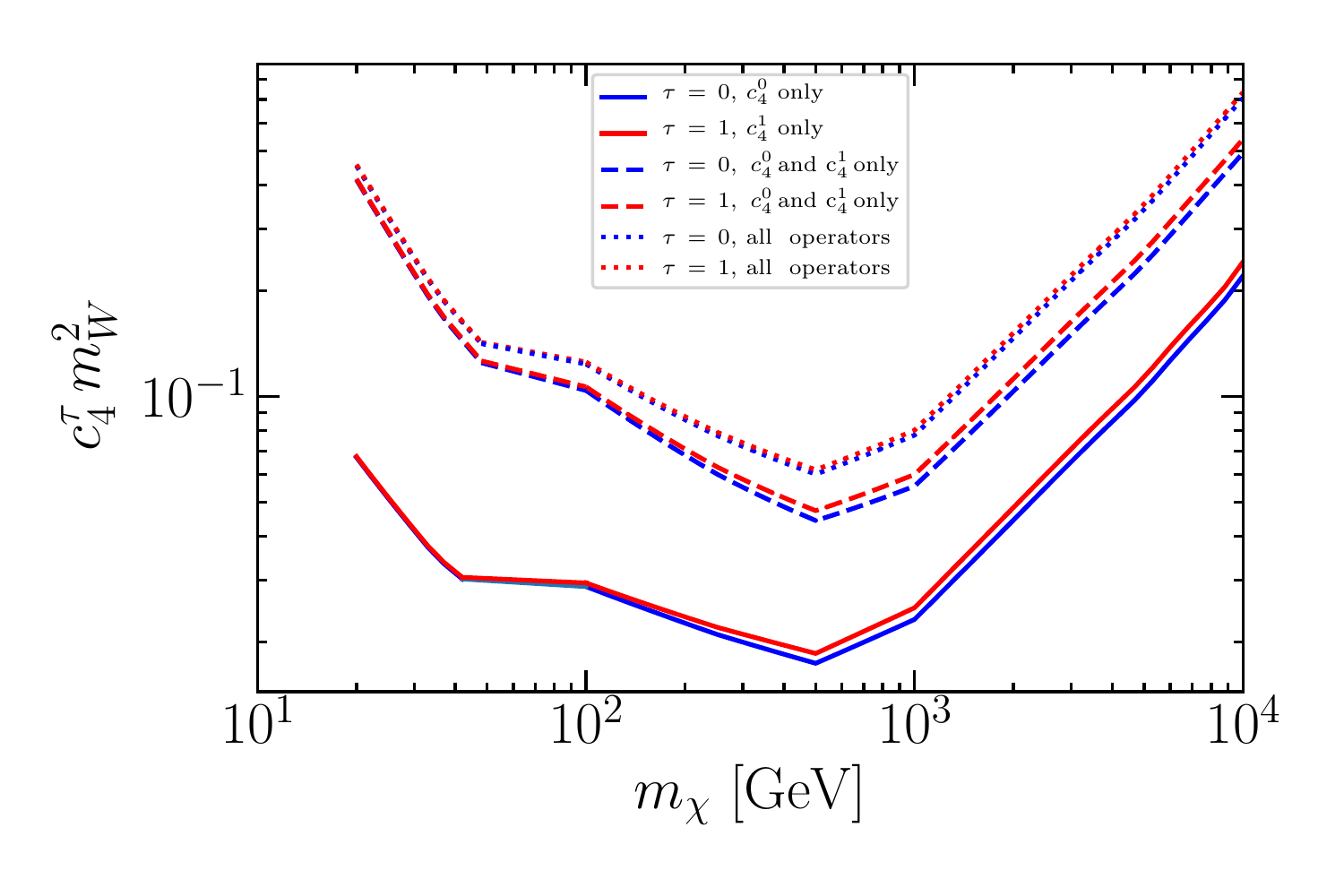}
		\includegraphics[width=.45\textwidth]{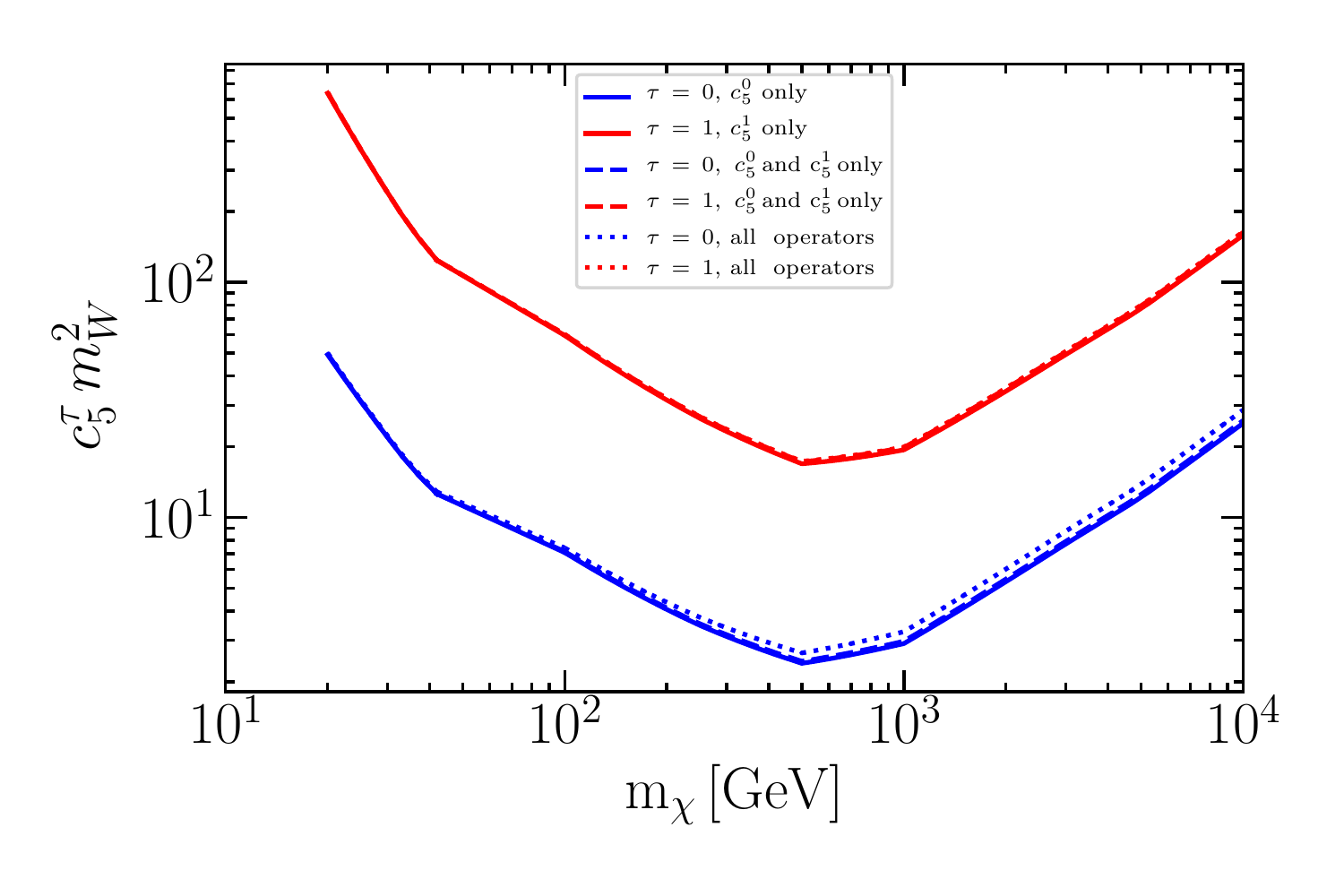}\\
		\includegraphics[width=.45\textwidth]{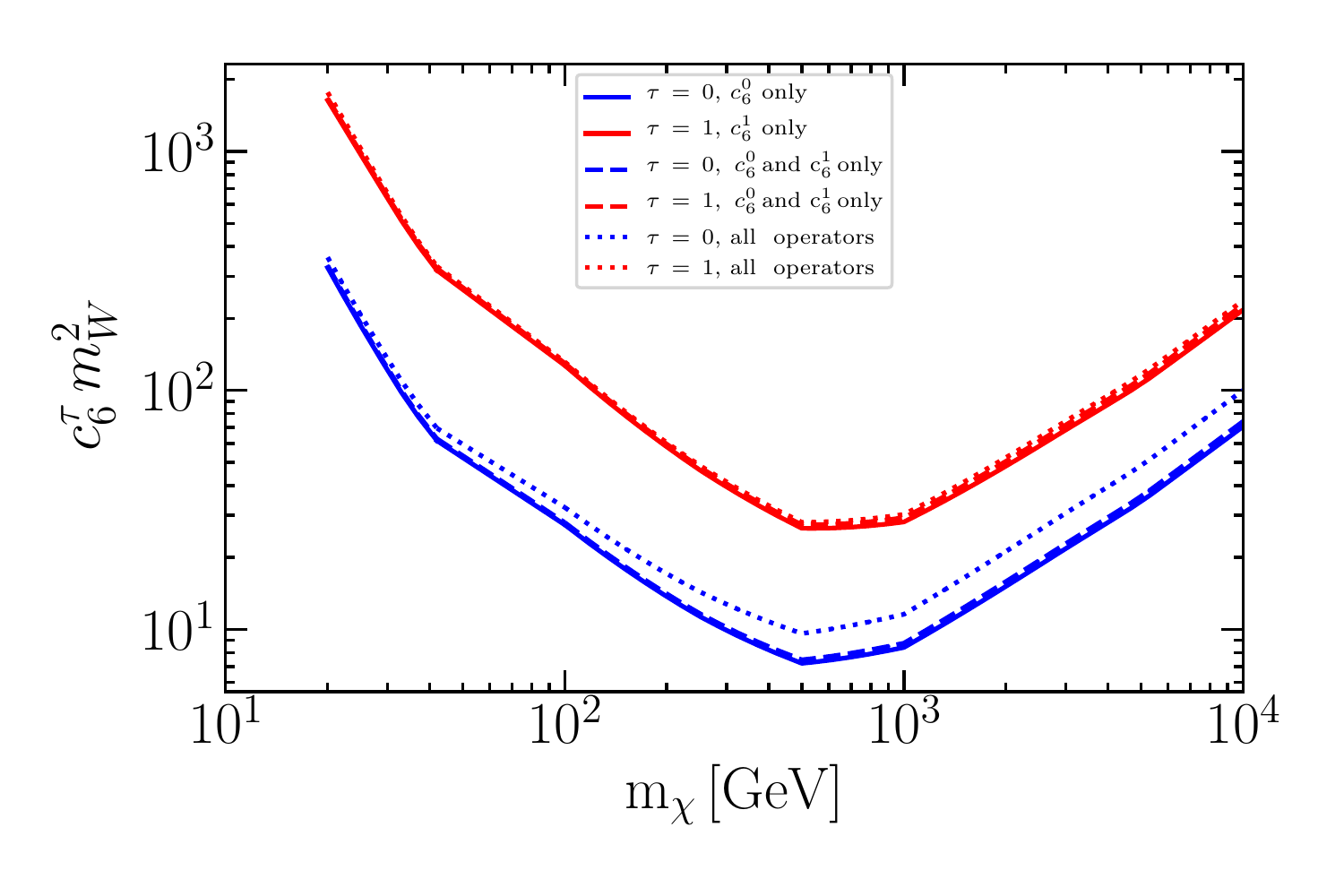}
		\includegraphics[width=.45\textwidth]{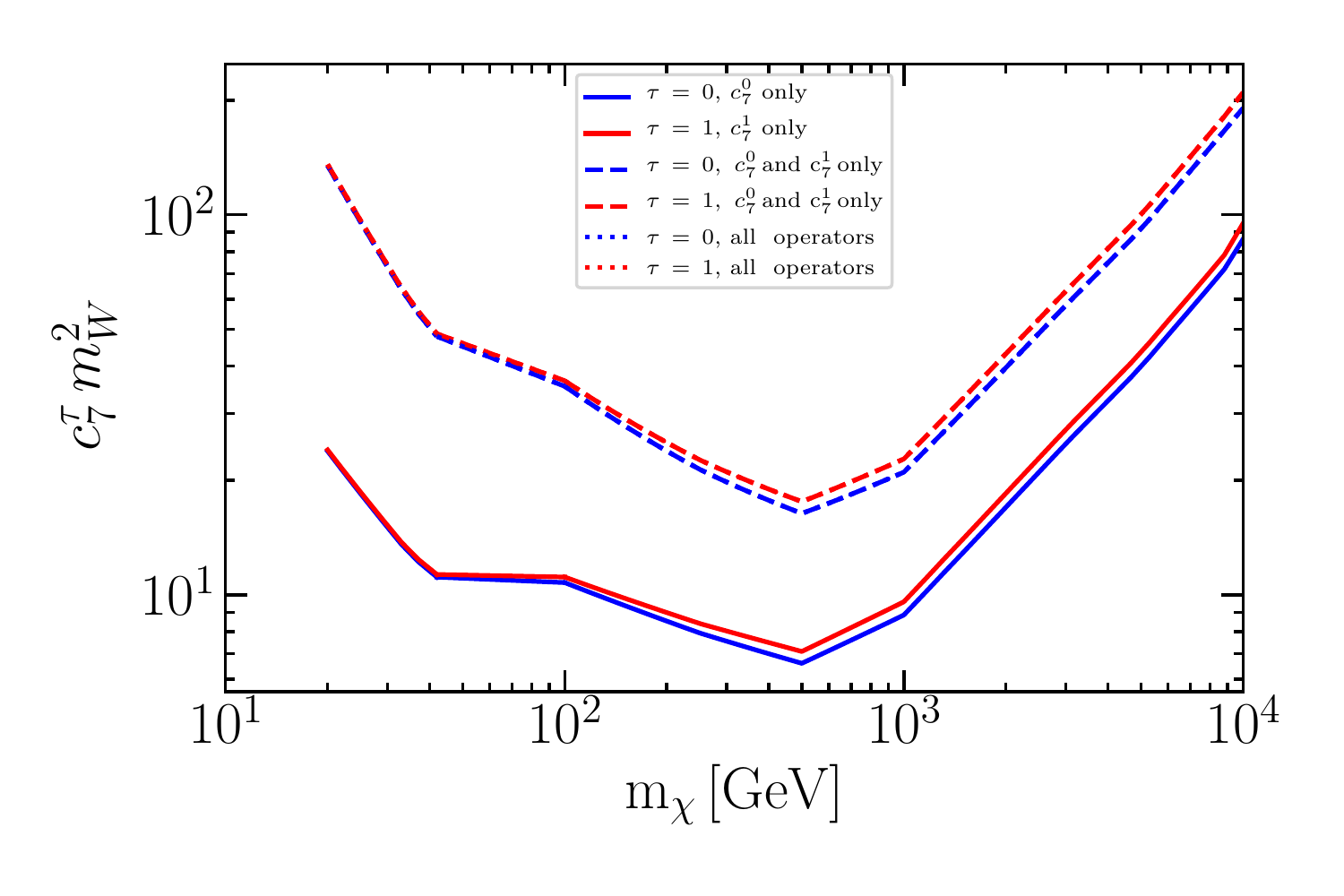}\\
		\includegraphics[width=.45\textwidth]{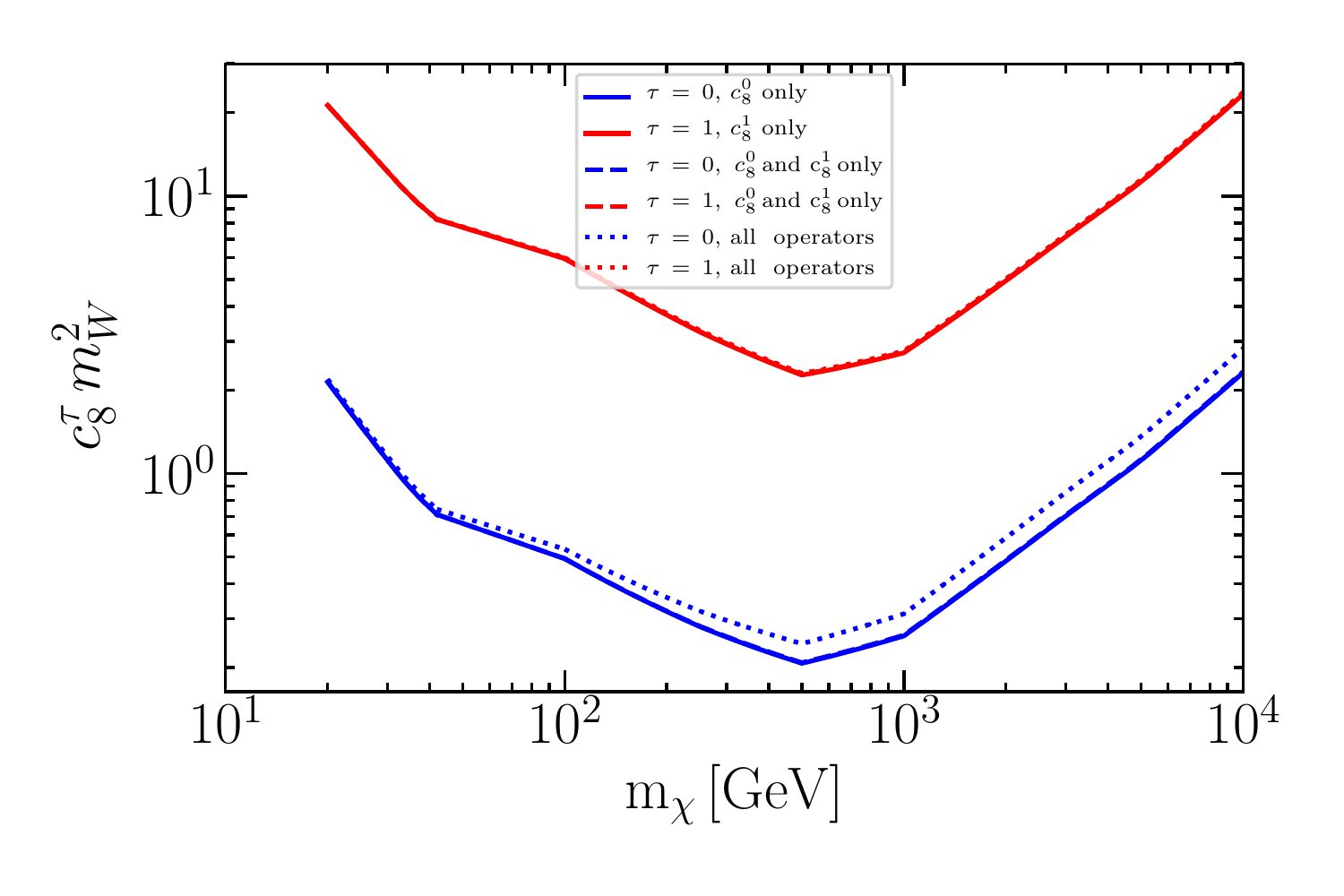}
		\includegraphics[width=.45\textwidth]{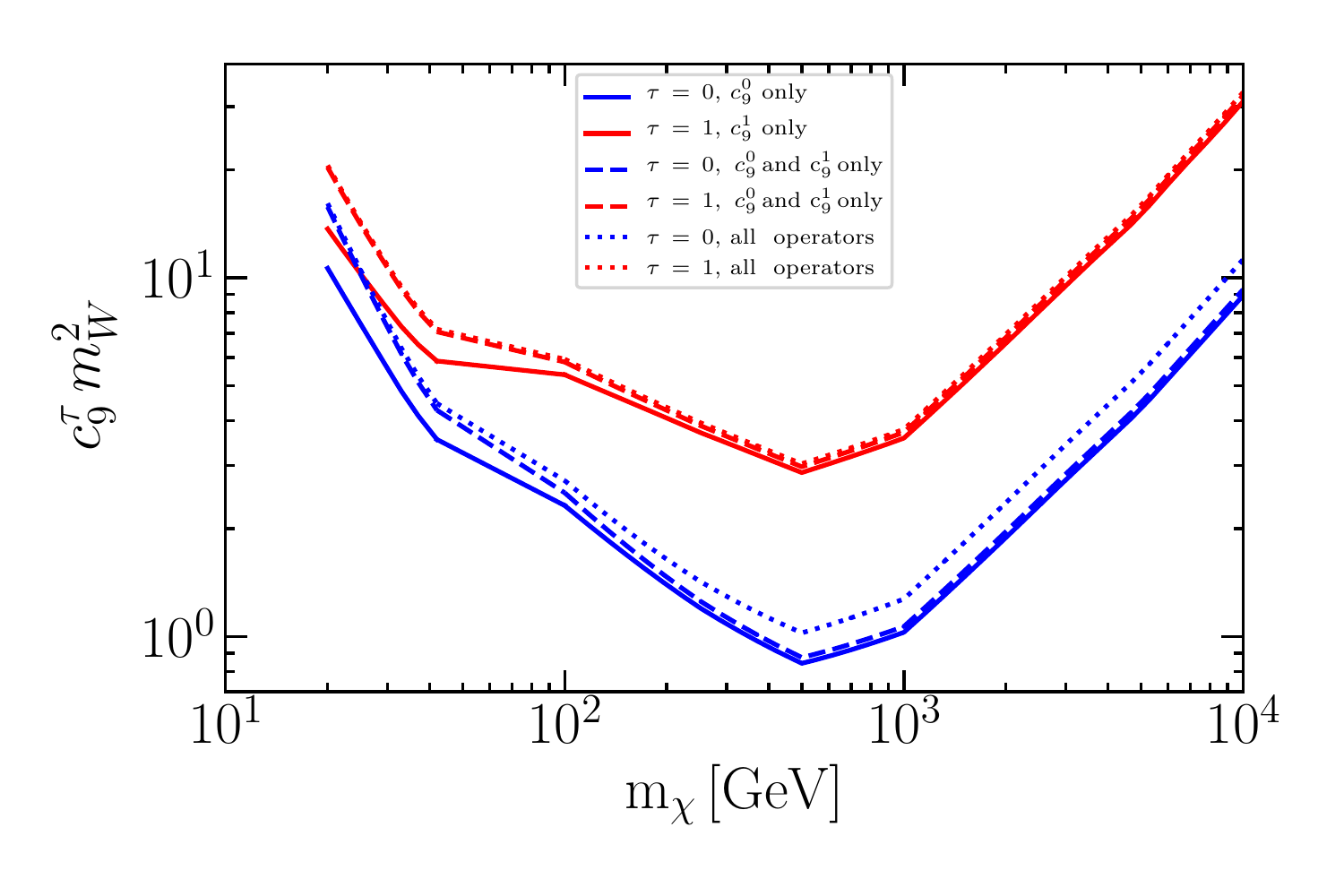}\\
	\end{center}
	\caption{\small Upper limits at the 90\% C.L. on the  coupling strengths of the isoscalar $c_i^0$ ( blue) and isovector $c_i^1$ (red) dark matter-nucleon interactions, $i=1...9$, for the annihilation final state $W^+W^-$ ($\tau^+\tau^-$ for $m_\chi<100$ GeV), considering for a given $i$ either the isoscalar or isovector interactions, {\it i.e.} the single operators $\widehat{\cal O}^0_i$ or $\widehat{\cal O}^1_i$  with no interference (solid); considering the interference of the isoscalar and isovector interactions, {\it i.e.}  the interference between the two operators $\widehat{\cal O}^0_i$ and $\widehat{\cal O}^1_i$ (dashed); and considering the interference of all isoscalar and isovector interactions, {\it i.e.} the interference of the 28 operators $\widehat{\cal O}^0_j$ and $\widehat{\cal O}^1_j$, $j=1...14$ (dotted).
}
	\label{fig:UL-c-page1}
\end{figure}

\begin{figure}[h!]
	\begin{center}
		\includegraphics[width=.45\textwidth]{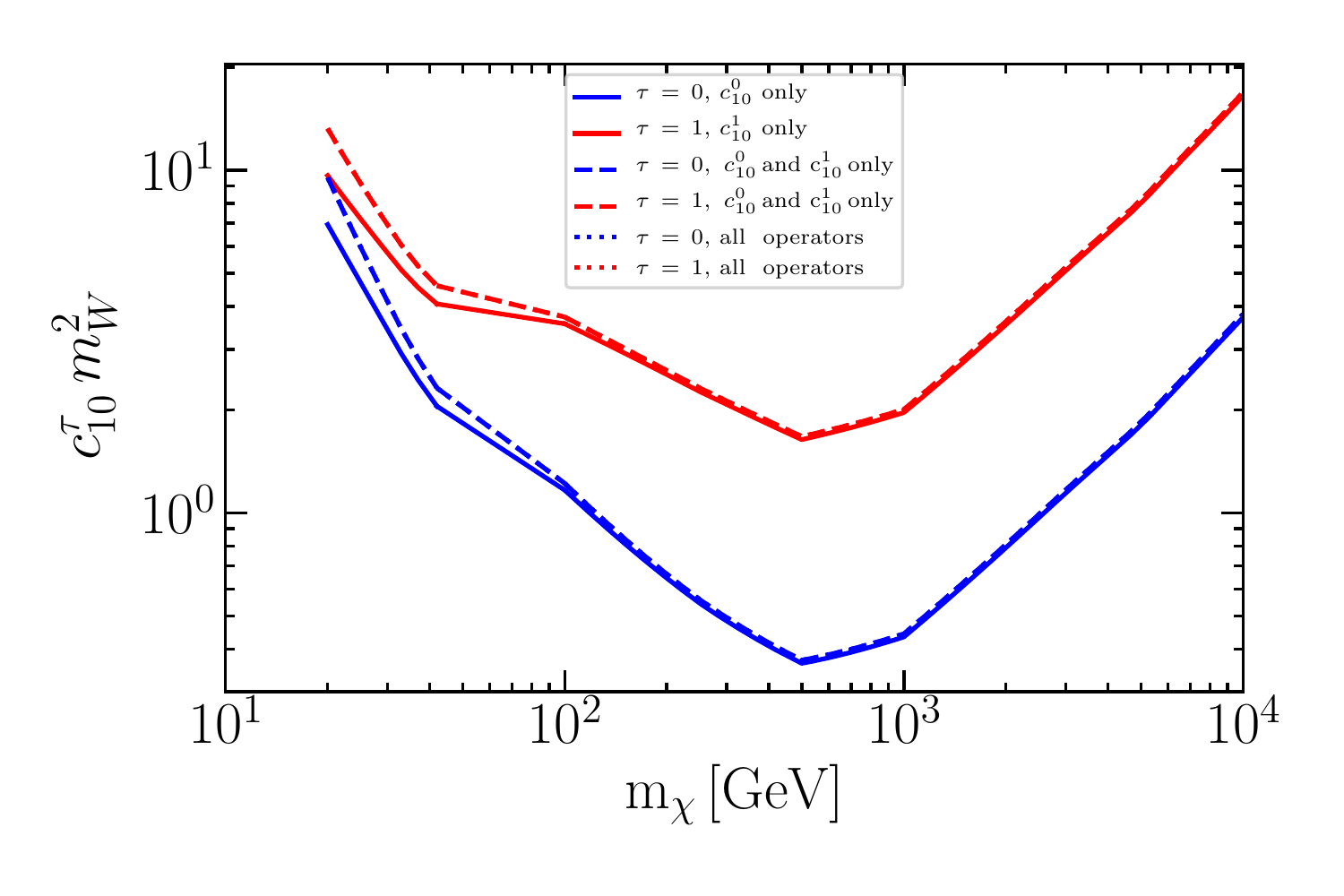}
		\includegraphics[width=.45\textwidth]{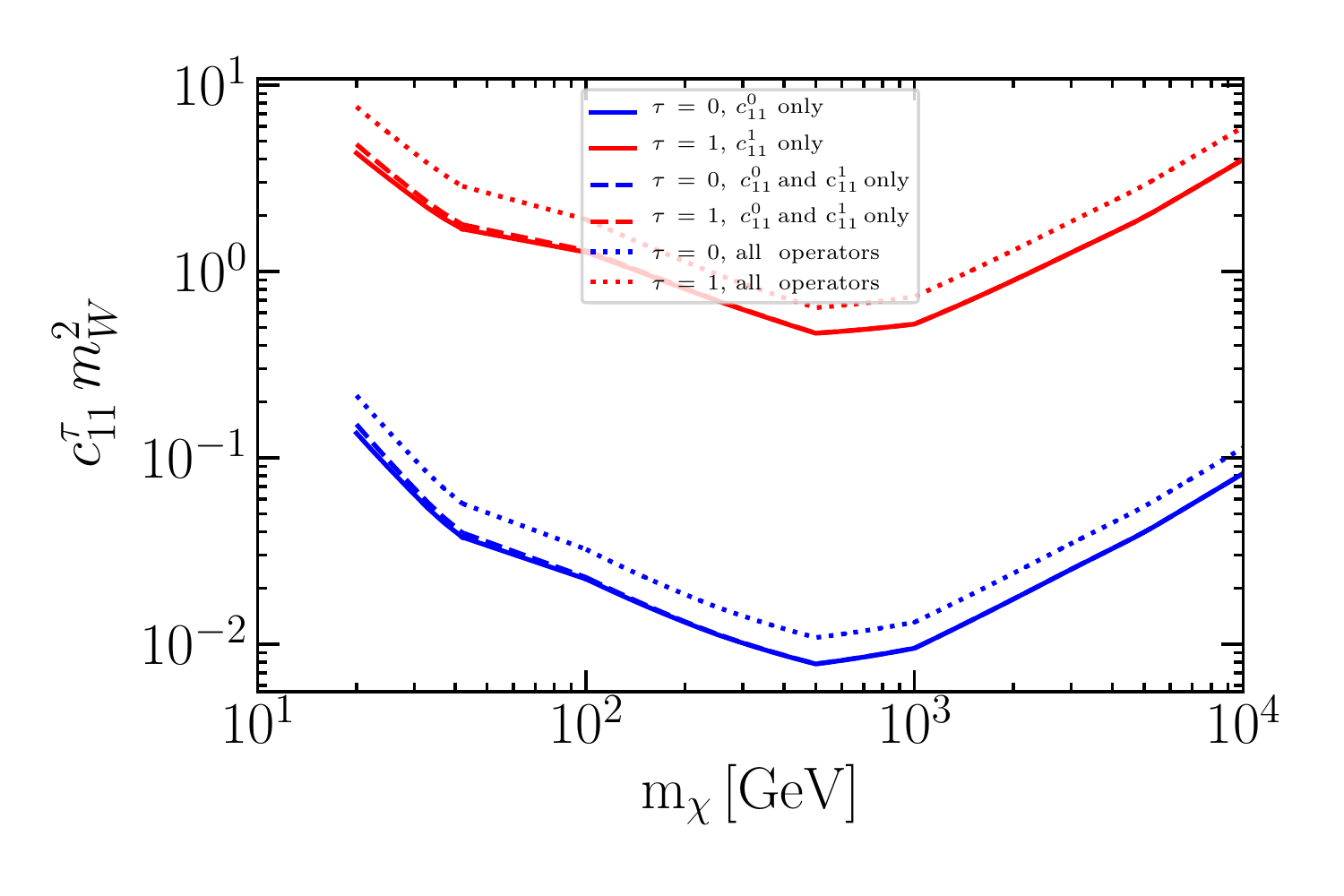}\\
		\includegraphics[width=.45\textwidth]{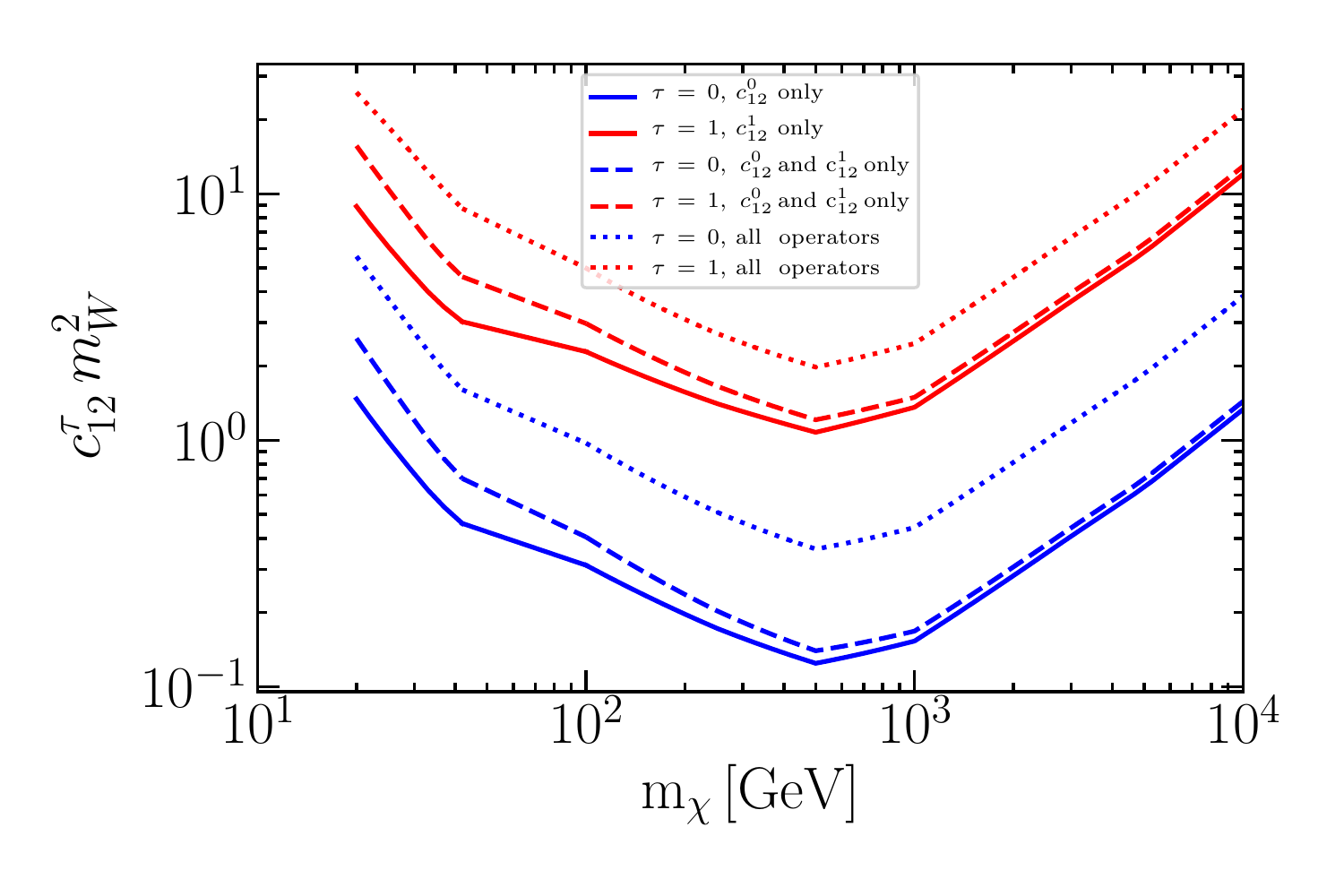}
		\includegraphics[width=.45\textwidth]{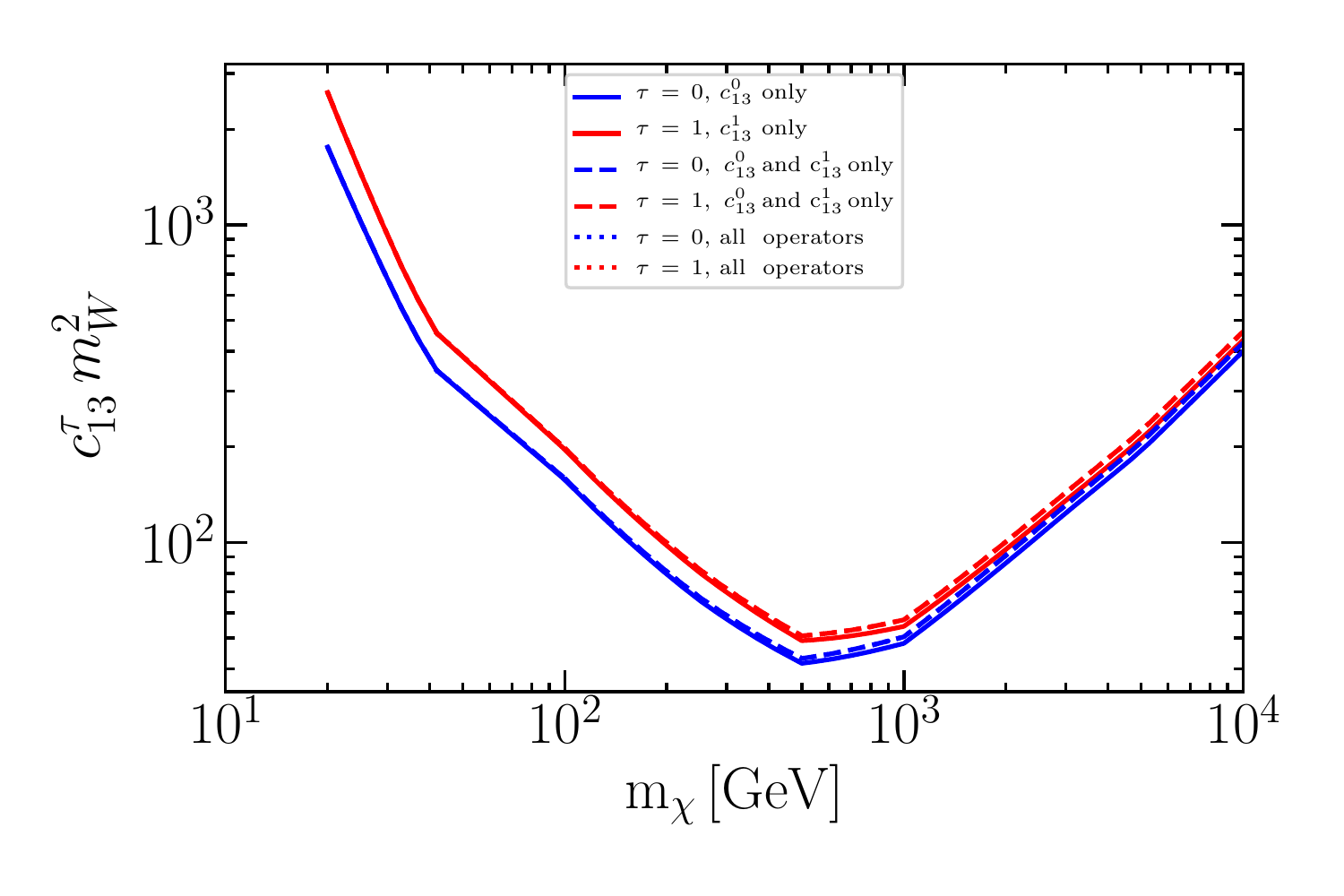}\\
		\includegraphics[width=.45\textwidth]{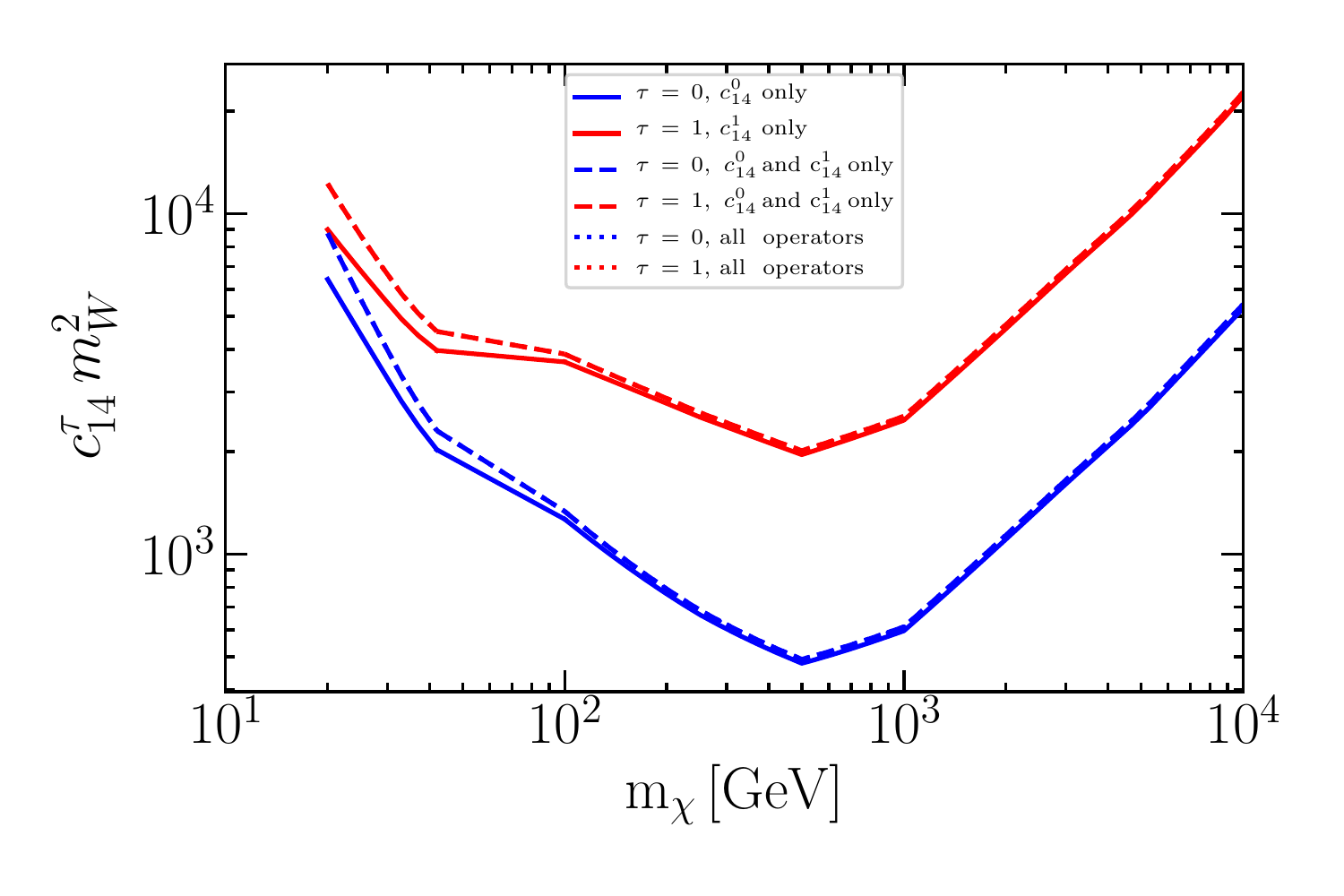}
		\includegraphics[width=.45\textwidth]{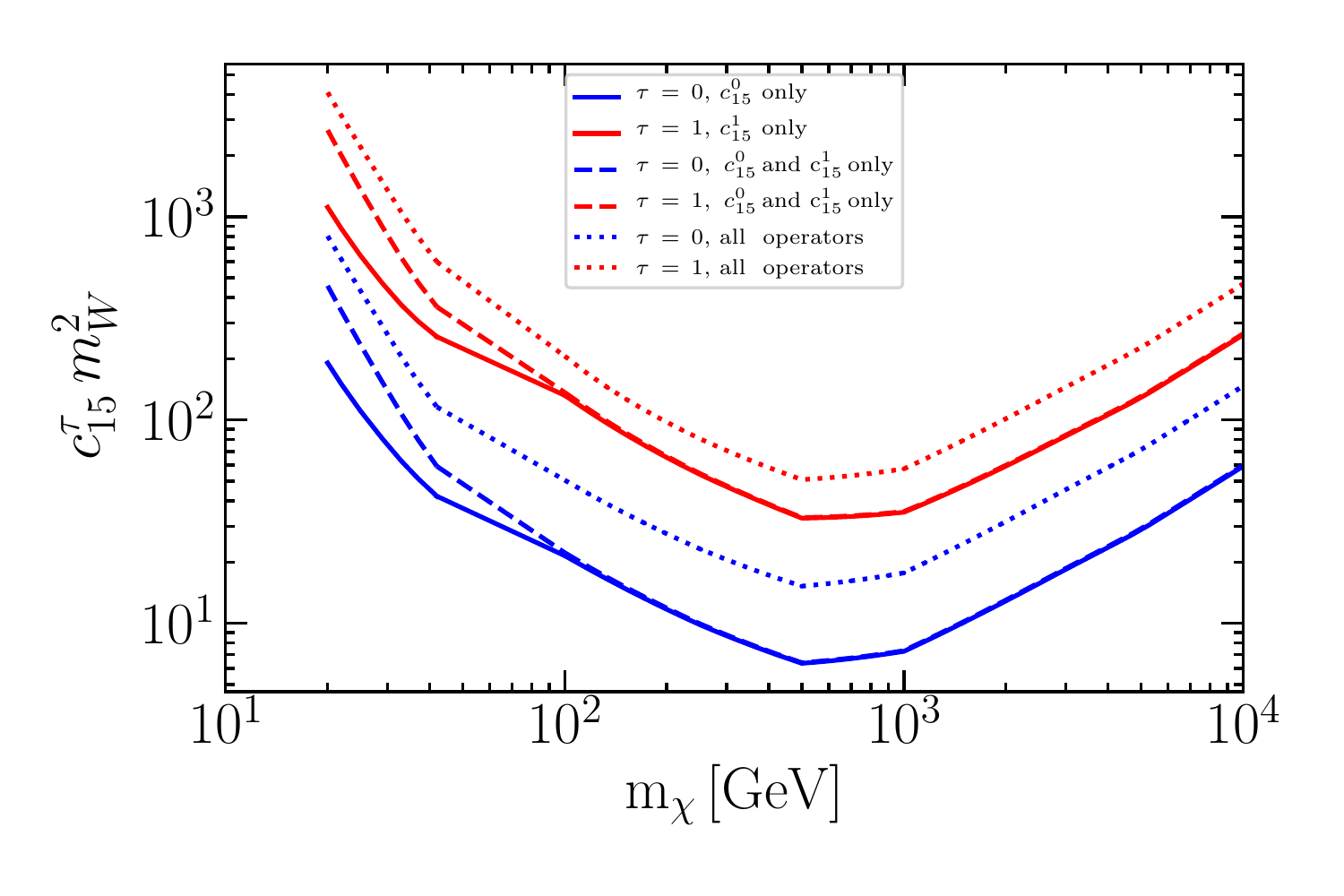}
	\end{center}
	\caption{\small Same as Fig. \ref{fig:UL-c-page1}, but for the operators $\widehat{\cal O}_i$, $i=8...15$.}
	\label{fig:UL-c-page2}
\end{figure}

\begin{figure}[h!]
	\begin{center}
		\includegraphics[width=.49\textwidth]{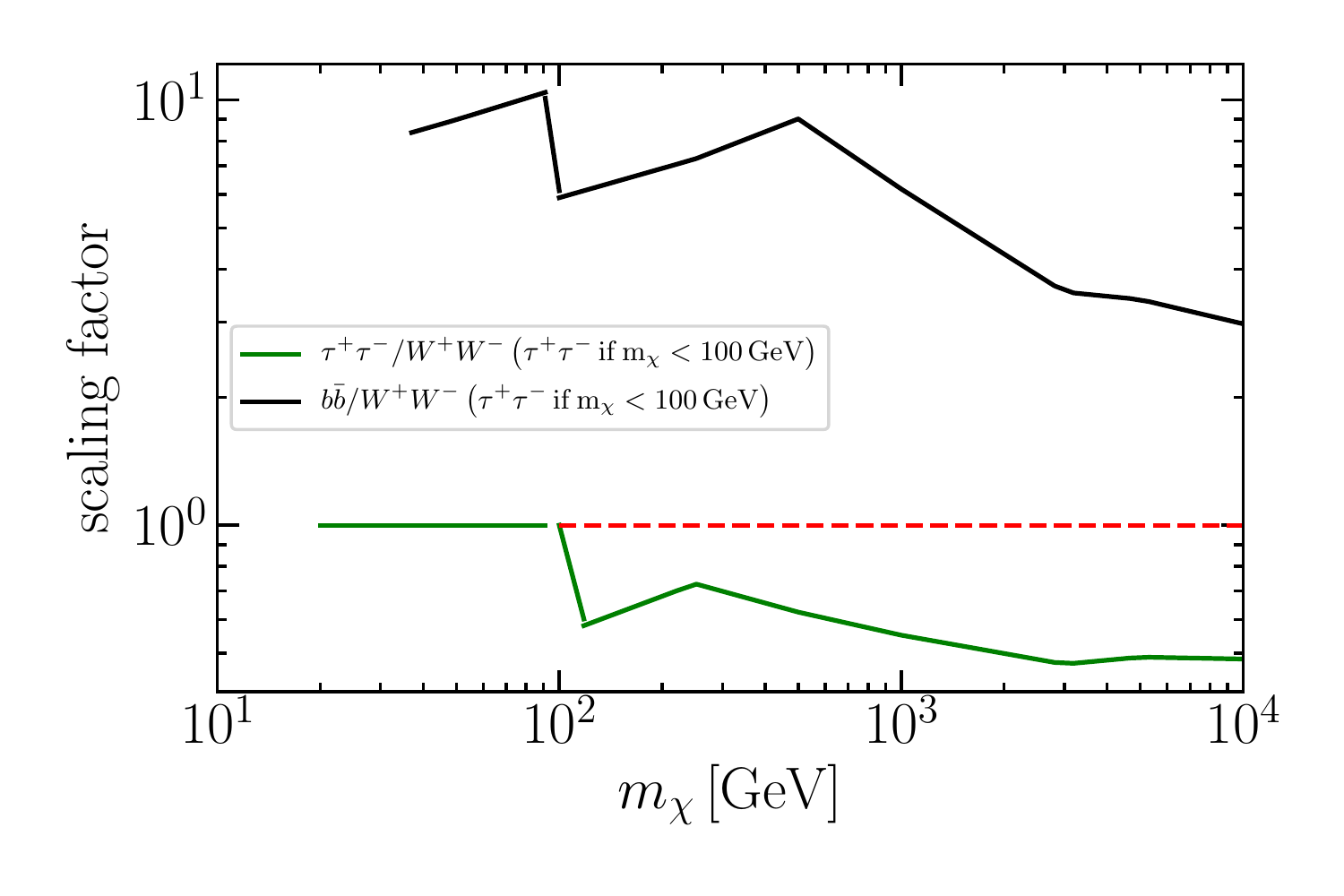}
	\end{center}
	\caption{\small Scaling factor to convert the limits on the coupling strengths in Figs. \ref{fig:UL-c-page1}, \ref{fig:UL-c-page2}, derived assuming annihilations into $W^+W^-$ ($\tau^+\tau^-$ for $m_\chi< 100$ GeV), into limits for annihilations into $b\bar b$ or into annihilations into $\tau^+\tau^-$.}
	\label{fig:ratio}
\end{figure}

Using the procedure described in Section \ref{sec:method} we have derived conservative 90\% C.L. upper limits on the coupling constants for each effective interaction. The results for the isoscalar-isovector basis are shown in Figs.~\ref{fig:UL-c-page1} and \ref{fig:UL-c-page2}, assuming annihilations into $W^+W^-$ (or into $\tau^+\tau^-$ for $m_{\chi}<100$ GeV). The limits for the other annihilation channels ($\tau^+\tau^-$ for $m_{\chi}>100$ GeV and $b \bar b$) can be derived from our limits applying the scaling factors shown in Fig.~\ref{fig:ratio}, which follow from the fact that under the assumption of equilibration between capture and annihilation, the upper limit on the annihilation rate translates into an upper limit on the capture rate that just depends on the annihilation channel and not on the nature of the underlying operator inducing the capture.

For each coupling constant we show as a solid blue (red) line the upper limit on $c_i^0$ ($c_i^1$) assuming that this is the only non-zero interaction; as a dashed blue (red) line the upper limit $c_i^0$ ($c_i^1$) assuming that both $c_i^0$ and $c_i^1$ can be non-zero and can interfere, while all other coupling constants vanish; and as a dotted blue (red) line the upper limit $c_i^0$ ($c_i^1$) assuming that all interactions are present (interference occurs between $\widehat{\mathcal{O}}_1$ and $\widehat{\mathcal{O}}_3$, between $\widehat{\mathcal{O}}_4$, $\widehat{\mathcal{O}}_5$ and $\widehat{\mathcal{O}}_6$, between   $\widehat{\mathcal{O}}_8$ and $\widehat{\mathcal{O}}_9$, and between $\widehat{\mathcal{O}}_{11}$, $\widehat{\mathcal{O}}_{12}$ and $\widehat{\mathcal{O}}_{15}$ \cite{Fitzpatrick:2012ix}). Clearly, the strongest limit corresponds to the case where only one interaction is present, and becomes weaker as interferences with more interactions become possible. It is important to stress that points in parameter space above the solid lines are not necessarily excluded. However, points above the dotted lines necessarily lead to a neutrino flux at Earth in conflict with the IceCube measurements (for annihilations into $W^+W^-$, or $\tau^+\tau^-$ for $m_{\chi}<100$ GeV, and assuming the Standard Halo Model).

The impact of the interference among interactions for deriving the upper limits on $c_i^\tau$, $\tau=0,1$, can be significant for some cases, such as for  $c_4^\tau$ or for $c_7^\tau$, and to a lesser extent for $c_{12}^\tau$ or for $c_{15}^\tau$. In contrast, it is rather modest for deriving the limits of other coupling strengths, such as for $c_{1}^\tau$, for $c_{5}^\tau$ or for $c_{13}^\tau$, for which the solid, dashed and dotted lines practically overlap. For the latter interactions,  the off-diagonal elements of the capture matrix $\mathbb{C}$ are, when expressed in the isoscalar-isovector basis, much smaller than the diagonal elements, so that the ellipse ${\bf c}^T \mathbb{C} {\bf c}=C^{\rm u.l.}$ is practically aligned to the axis; see Fig.~\ref{fig:ellipse_bases}.

\begin{figure}[h!]
	\begin{center}
		\includegraphics[width=.45\textwidth]{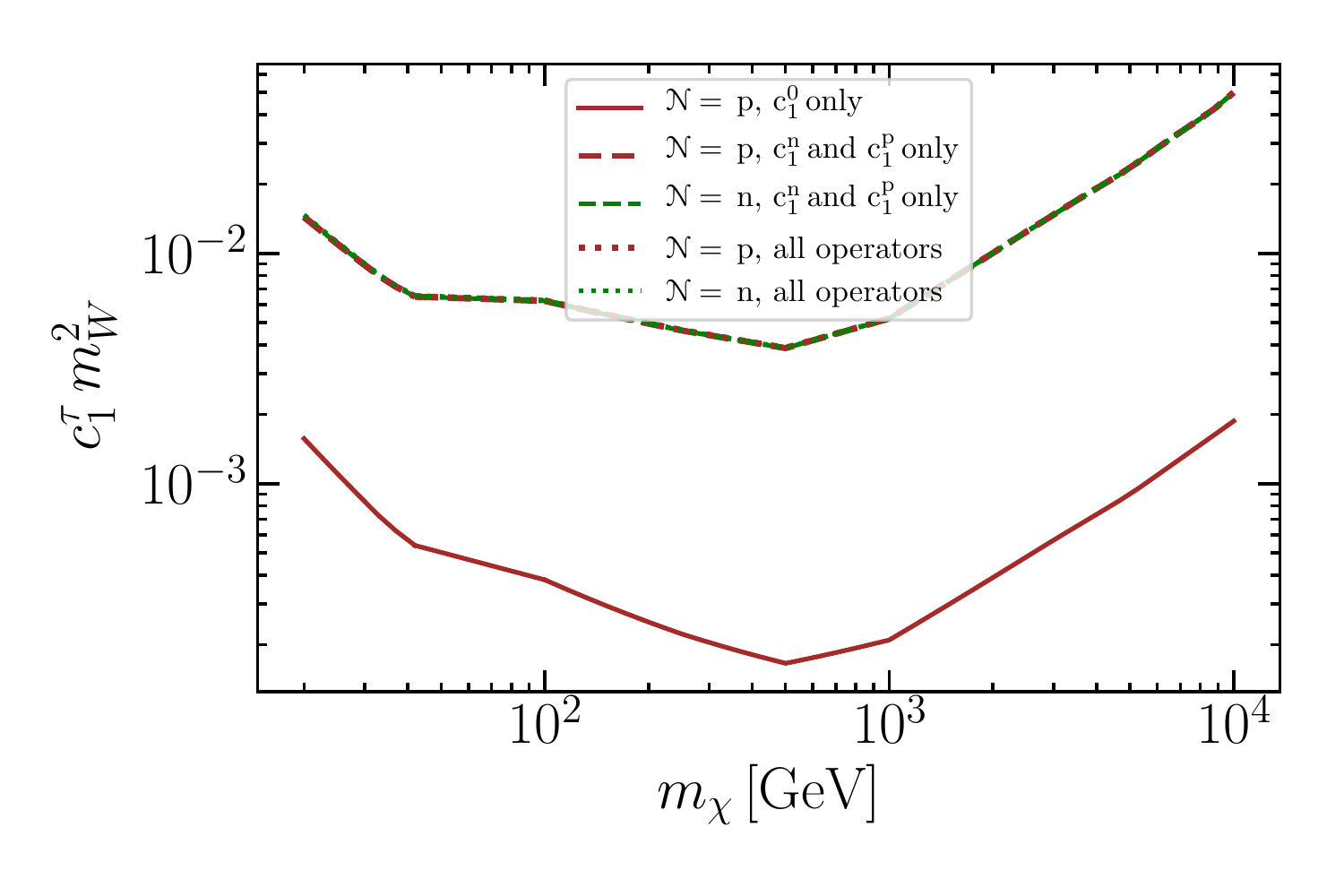}
		\includegraphics[width=.45\textwidth]{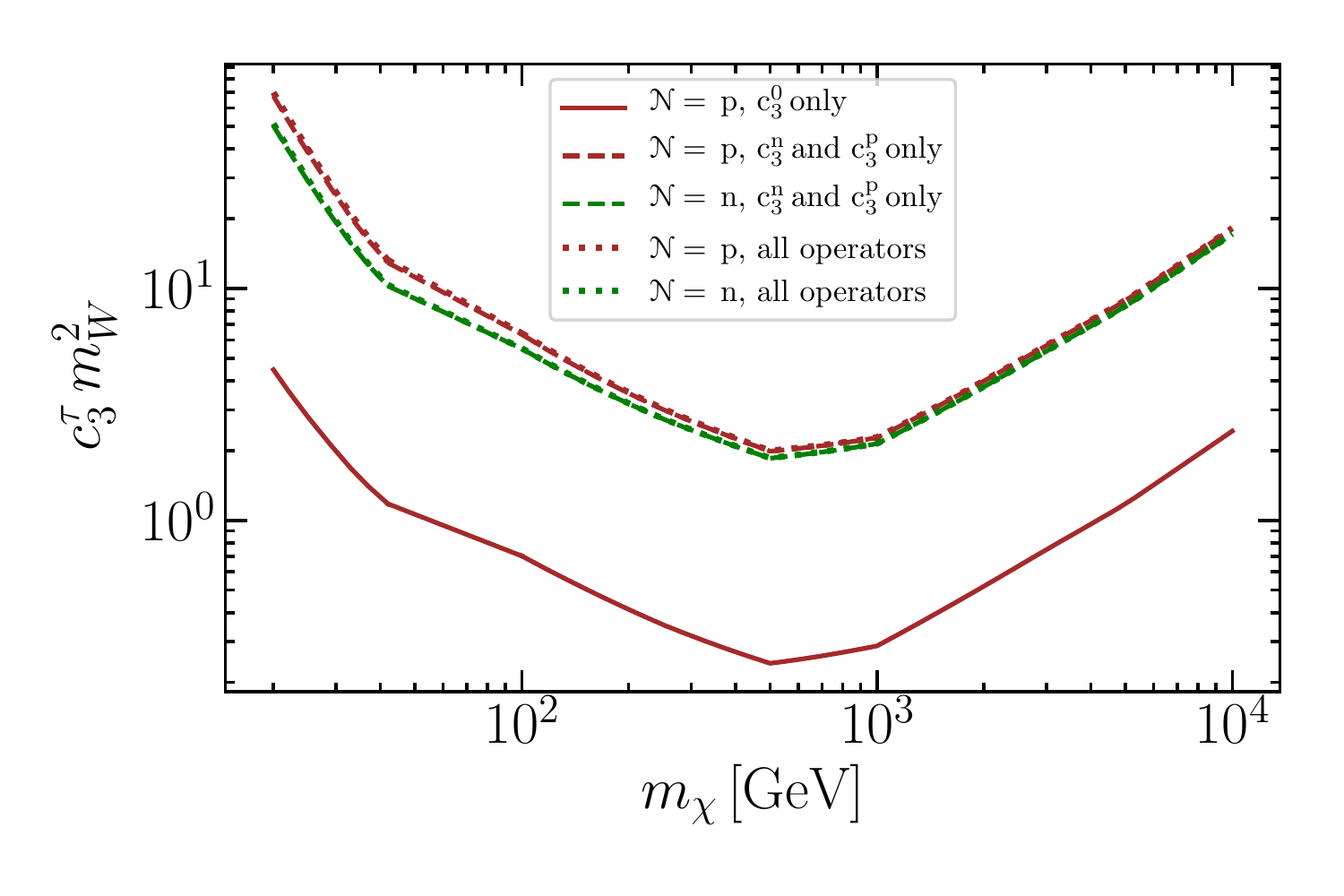}\\
		\includegraphics[width=.45\textwidth]{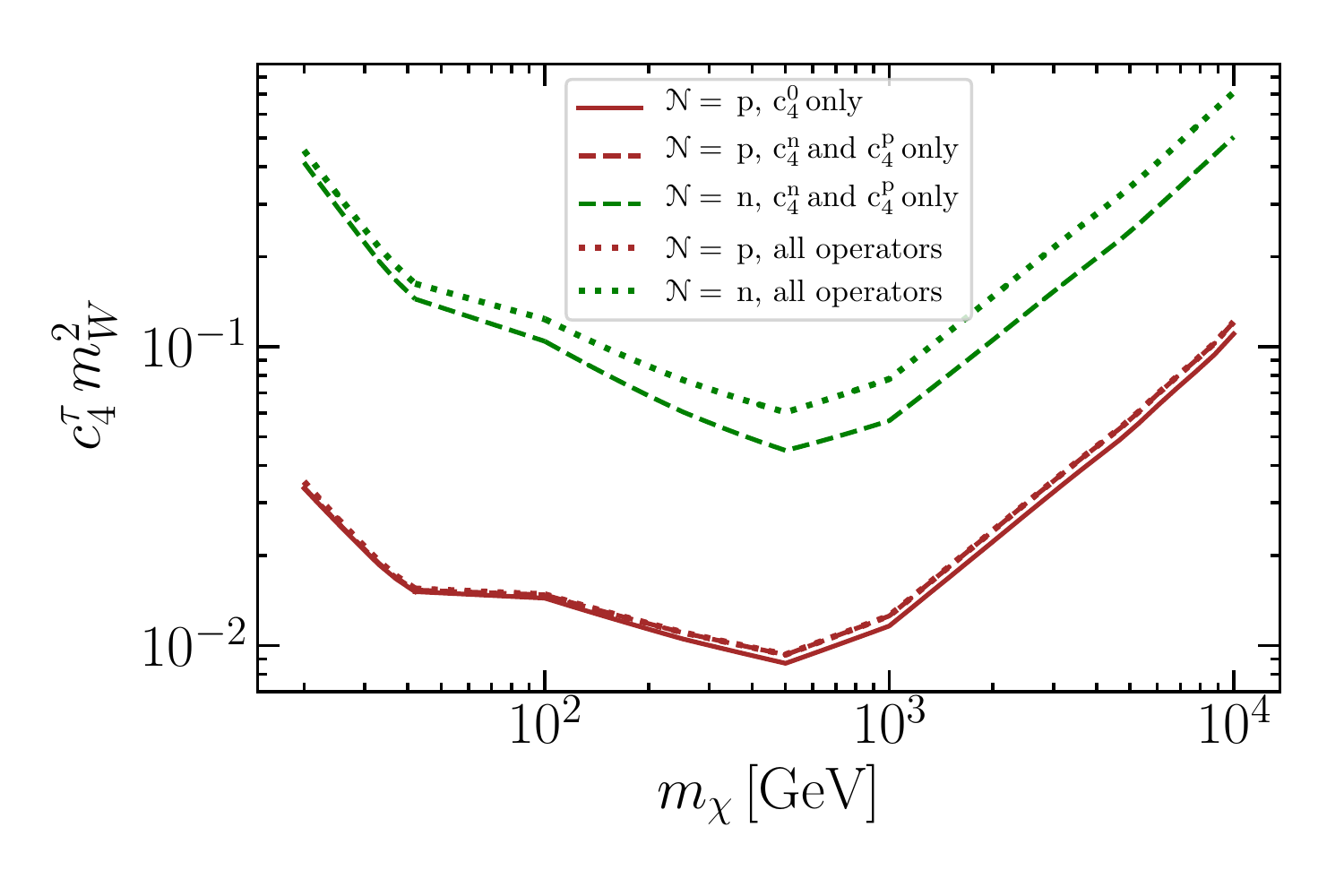}
		\includegraphics[width=.45\textwidth]{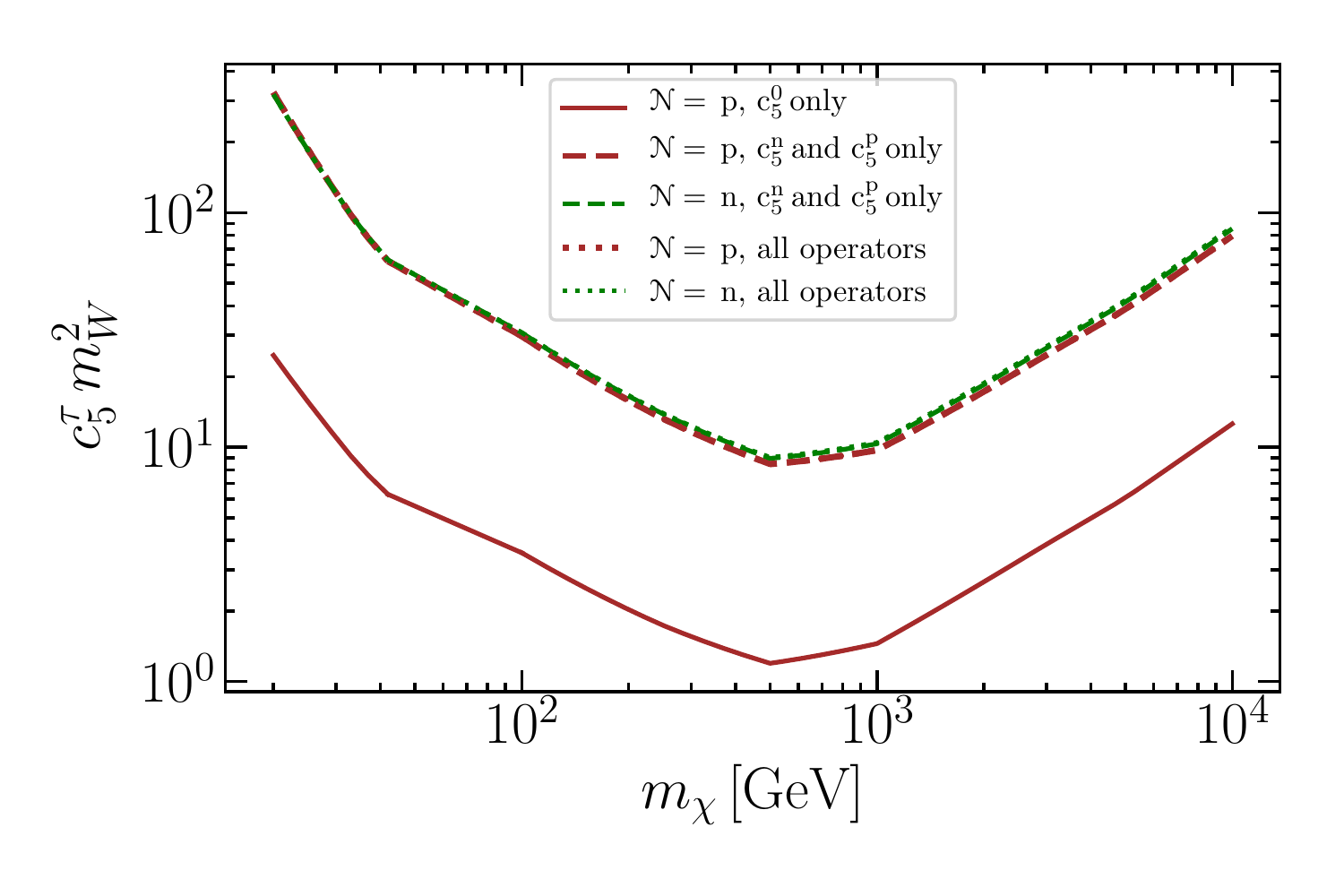}\\
		\includegraphics[width=.45\textwidth]{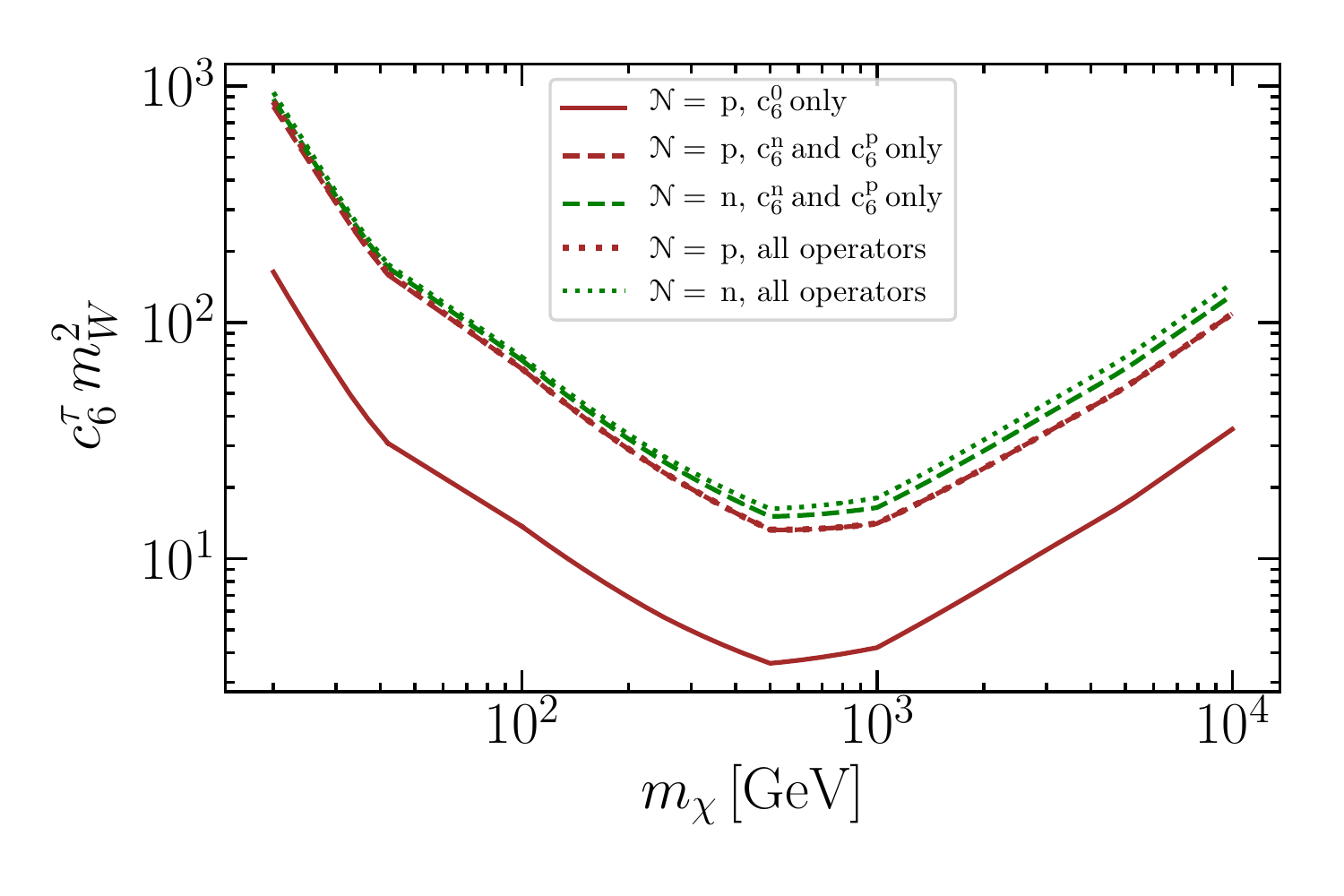}
		\includegraphics[width=.45\textwidth]{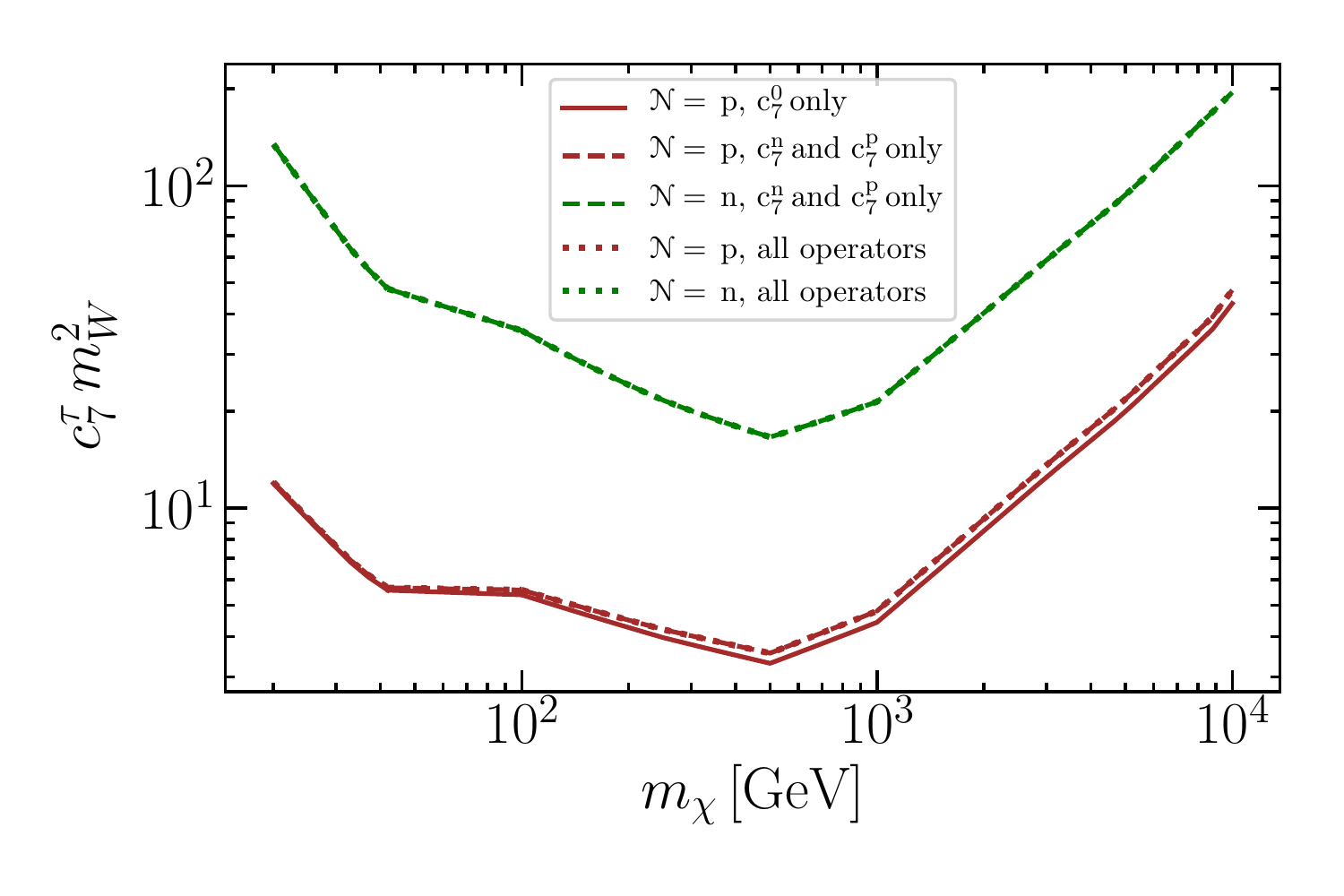}\\
		\includegraphics[width=.45\textwidth]{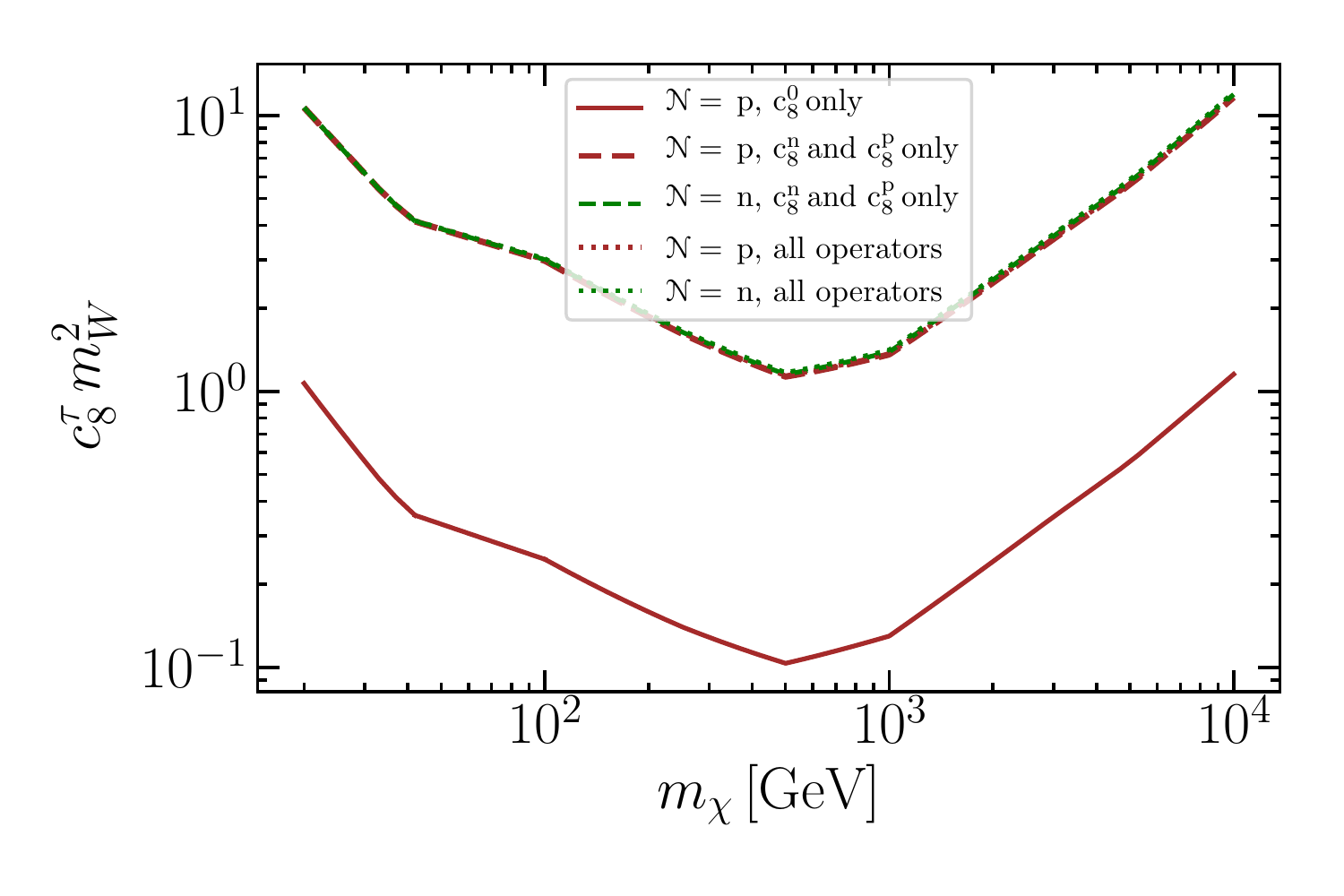}
		\includegraphics[width=.45\textwidth]{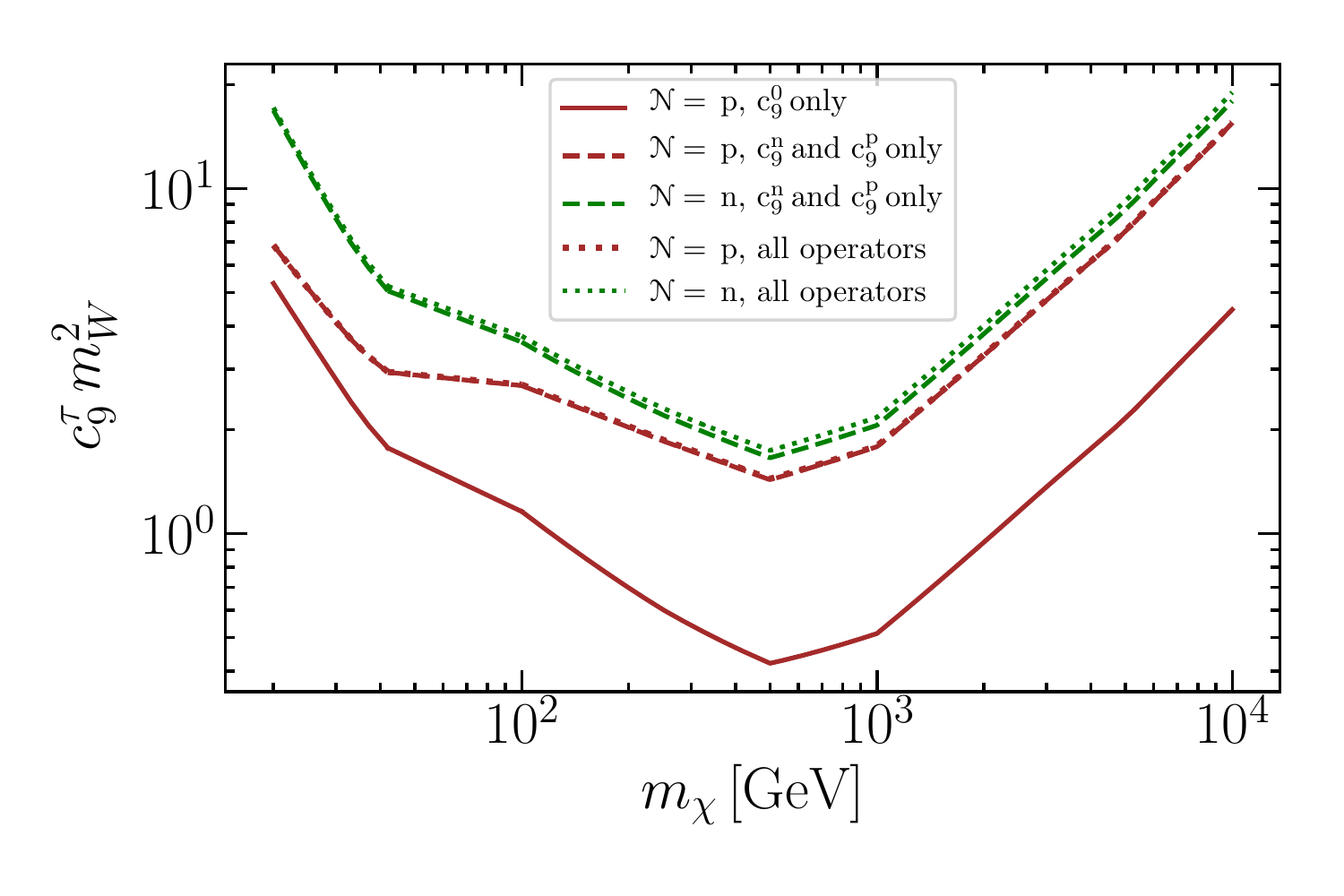}\\
	\end{center}
	\caption{\small Upper limits at the 90\% C.L. on the coupling strengths of the dark matter-proton $c_i^p$ (brown) and dark matter-neutron $c_i^n$(green), $i=1...9$, for the annihilation final state $W^+W^-$ ($\tau^+\tau^-$ for $m_\chi<100$ GeV), considering for a given $i$ only the isoscalar interactions, {\it i.e.} the single operators $\widehat{\cal O}^0_i$  with no interference (solid); considering the interference of the  neutron and proton interactions, {\it i.e.}  the interference between the two operators $\widehat{\cal O}^n_i$ and $\widehat{\cal O}^p_i$ (dashed); and considering the interference of all neutron and proton interactions, {\it i.e.} the interference of the 28 operators $\widehat{\cal O}^n_j$ and $\widehat{\cal O}^p_j$, $j=1...14$ (dotted).
}
	\label{fig:UL-cnp-page1}
\end{figure}

\begin{figure}[h!]
	\begin{center}
		\includegraphics[width=.45\textwidth]{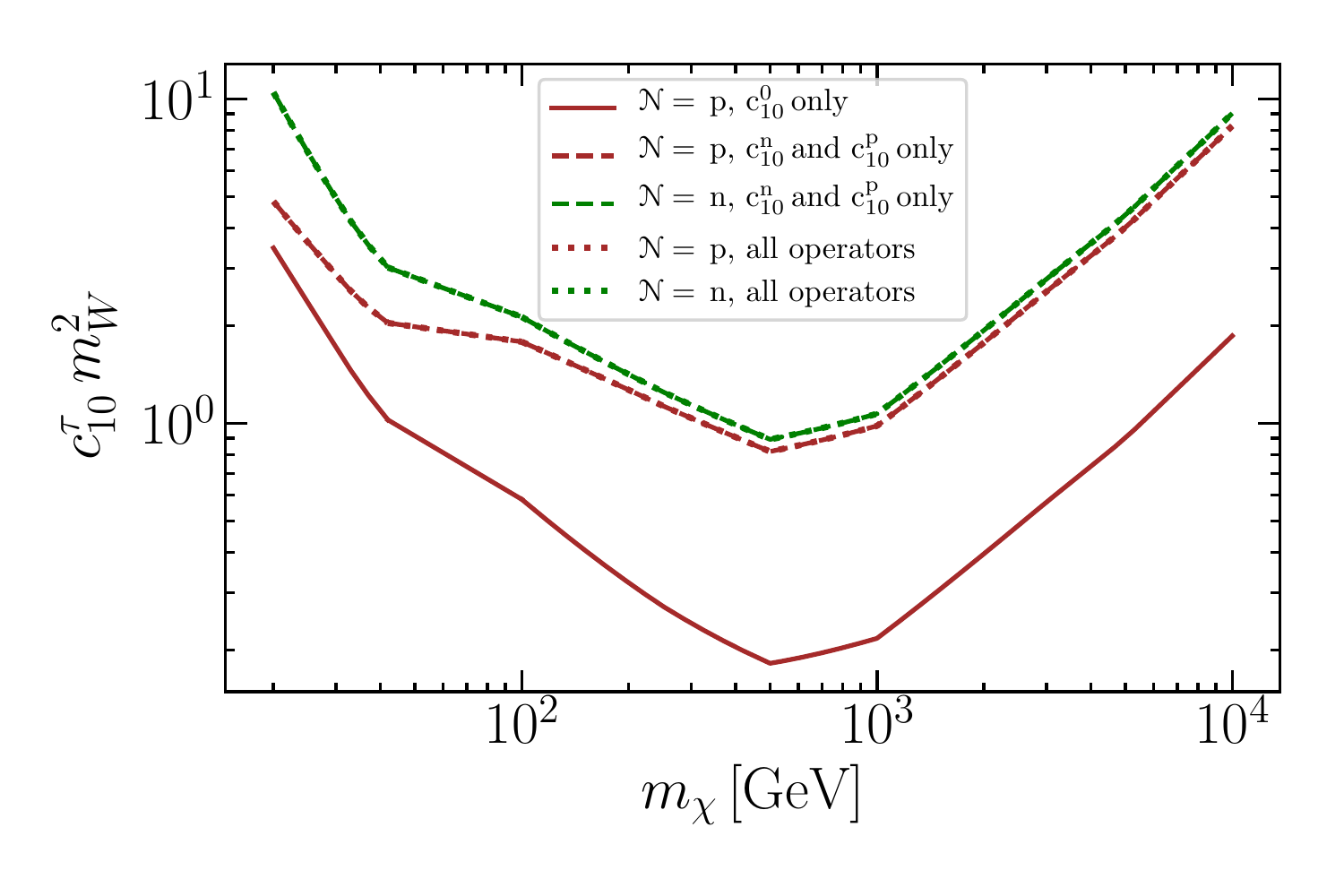}
		\includegraphics[width=.45\textwidth]{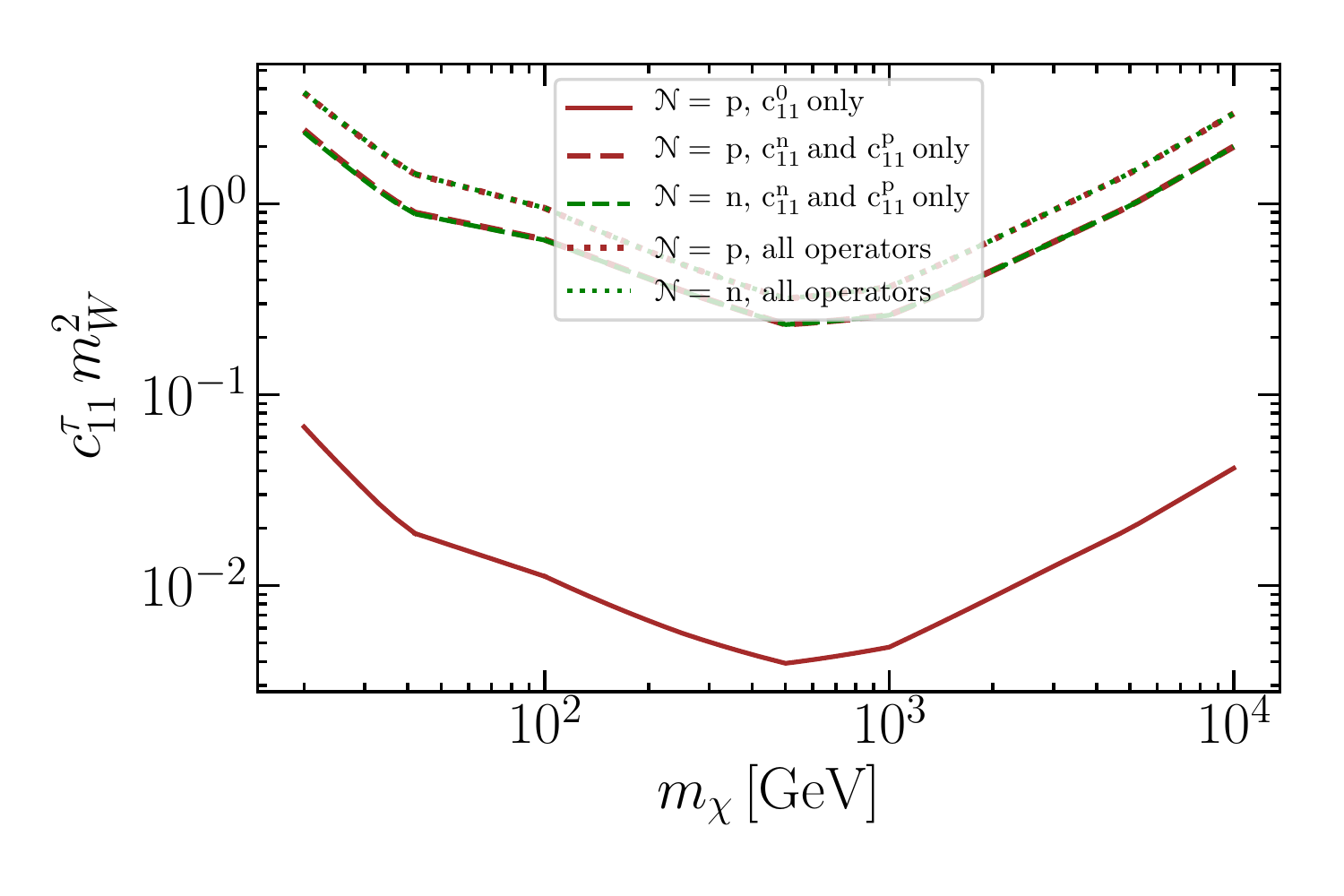}\\
		\includegraphics[width=.45\textwidth]{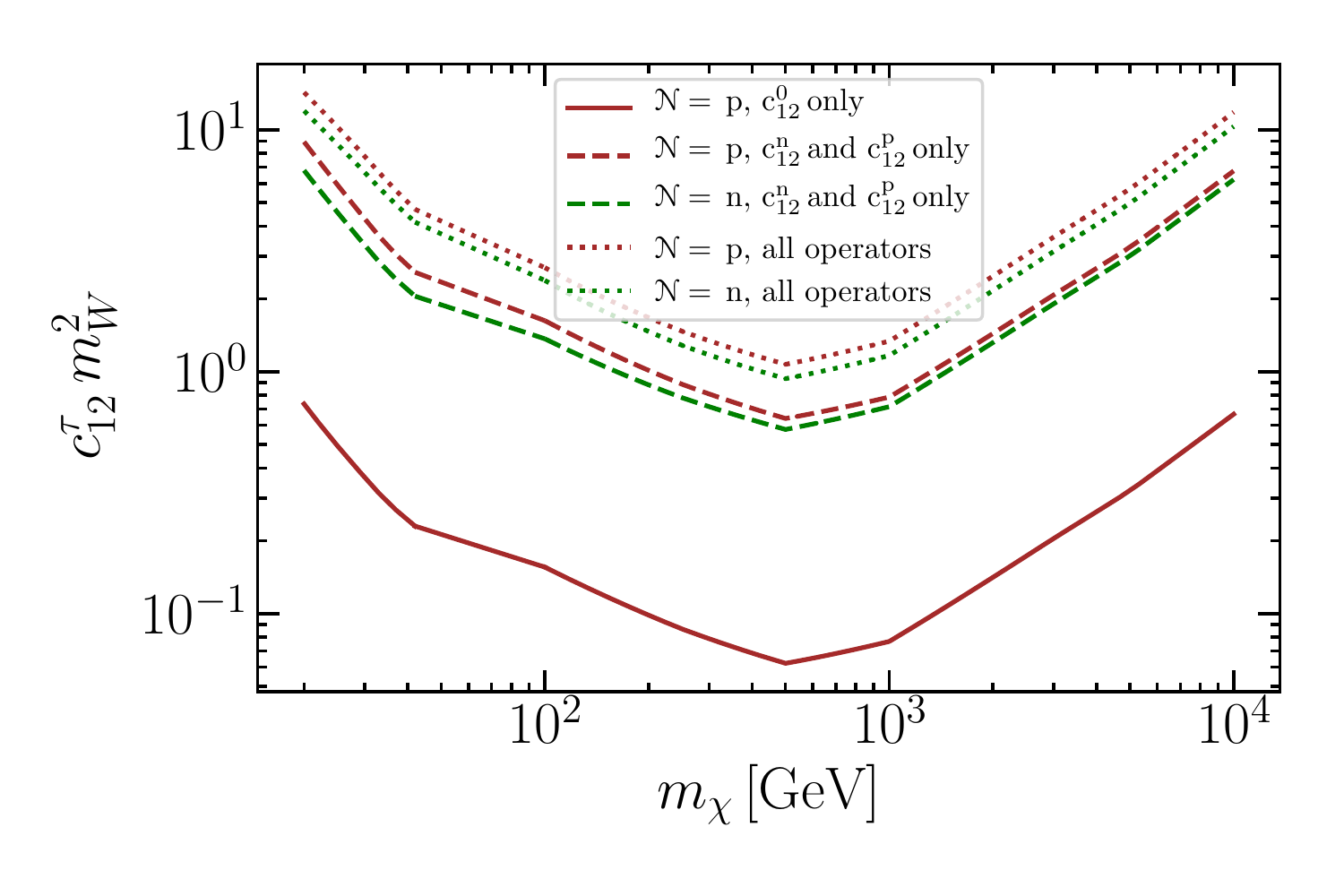}
		\includegraphics[width=.45\textwidth]{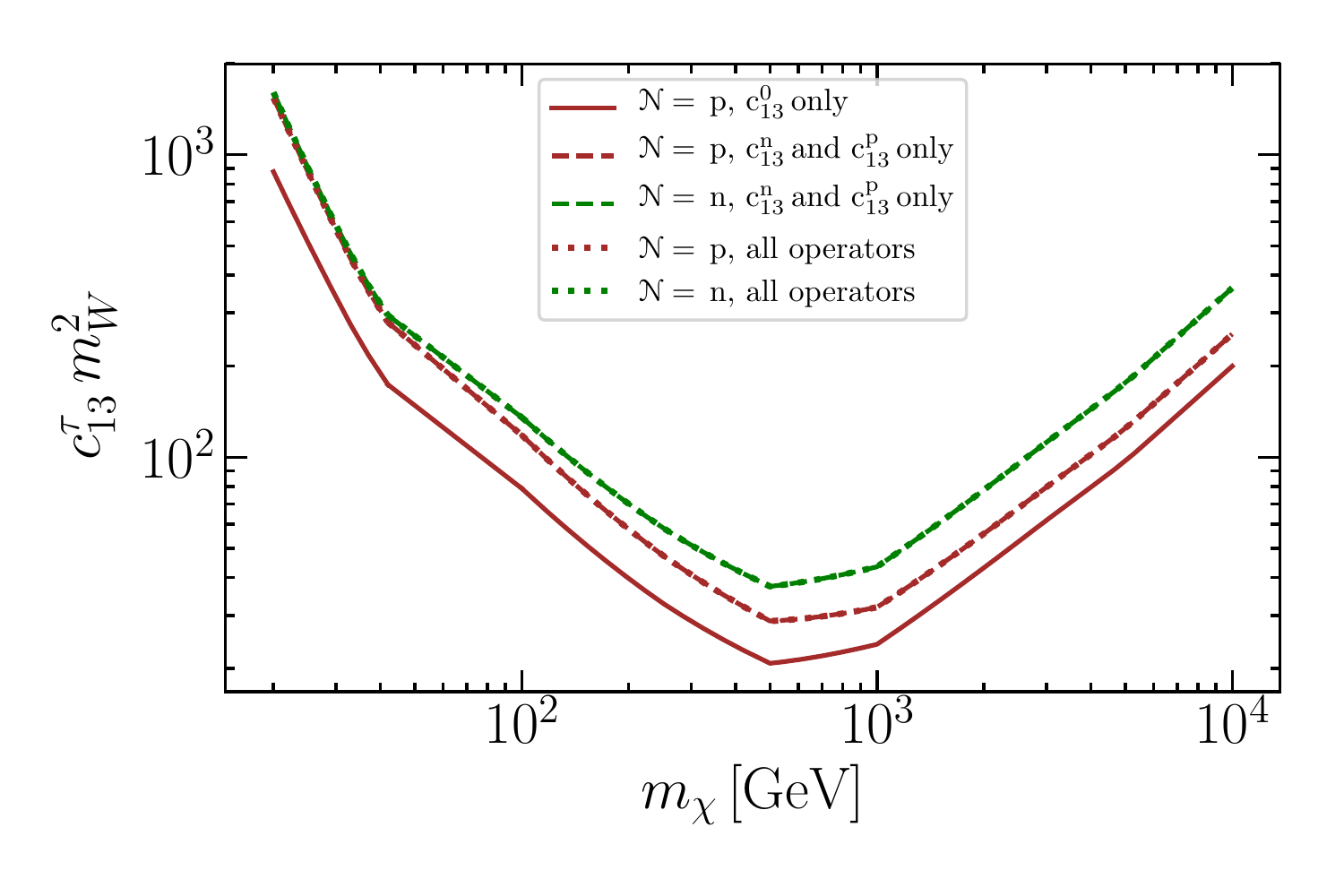}\\
		\includegraphics[width=.45\textwidth]{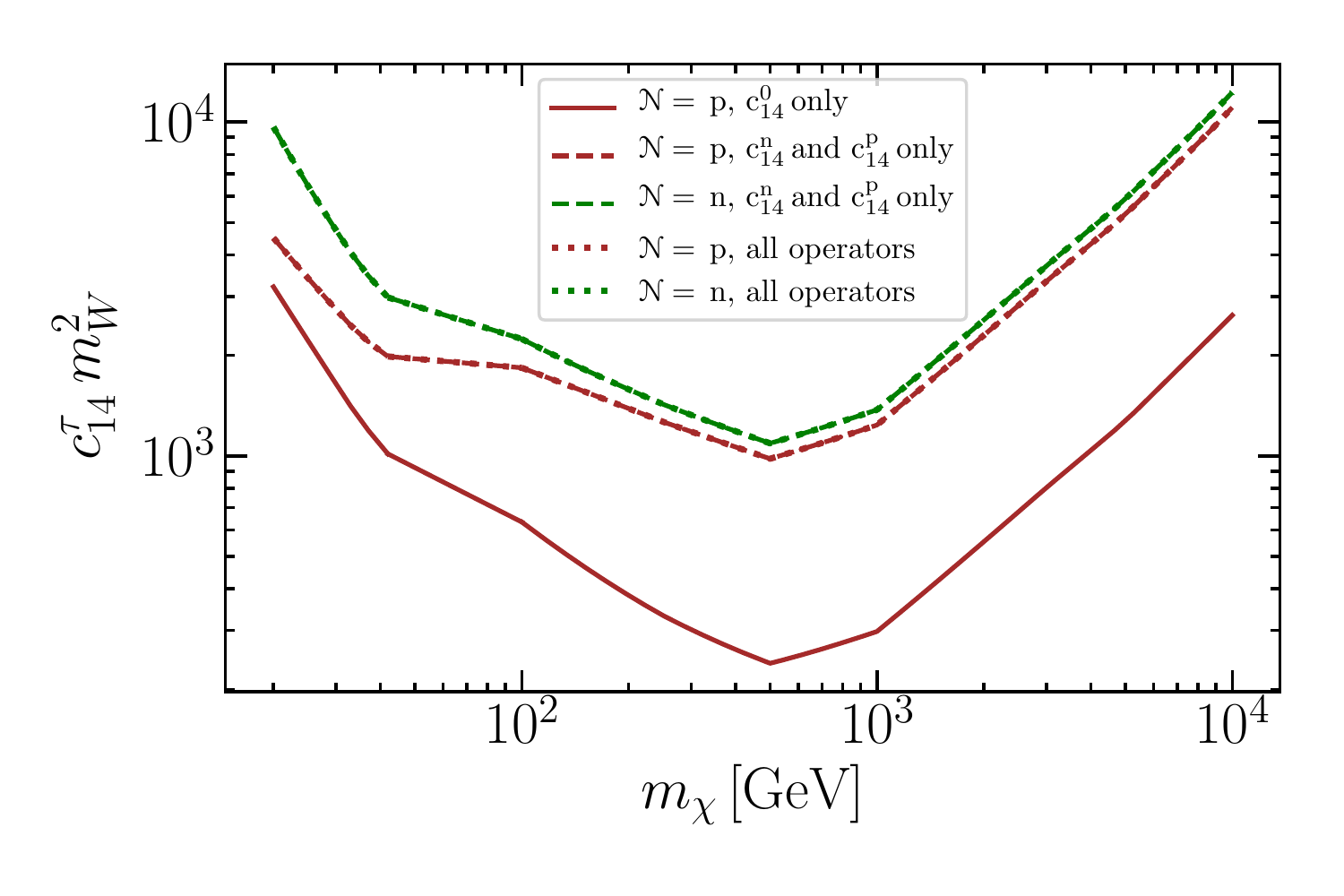}
		\includegraphics[width=.45\textwidth]{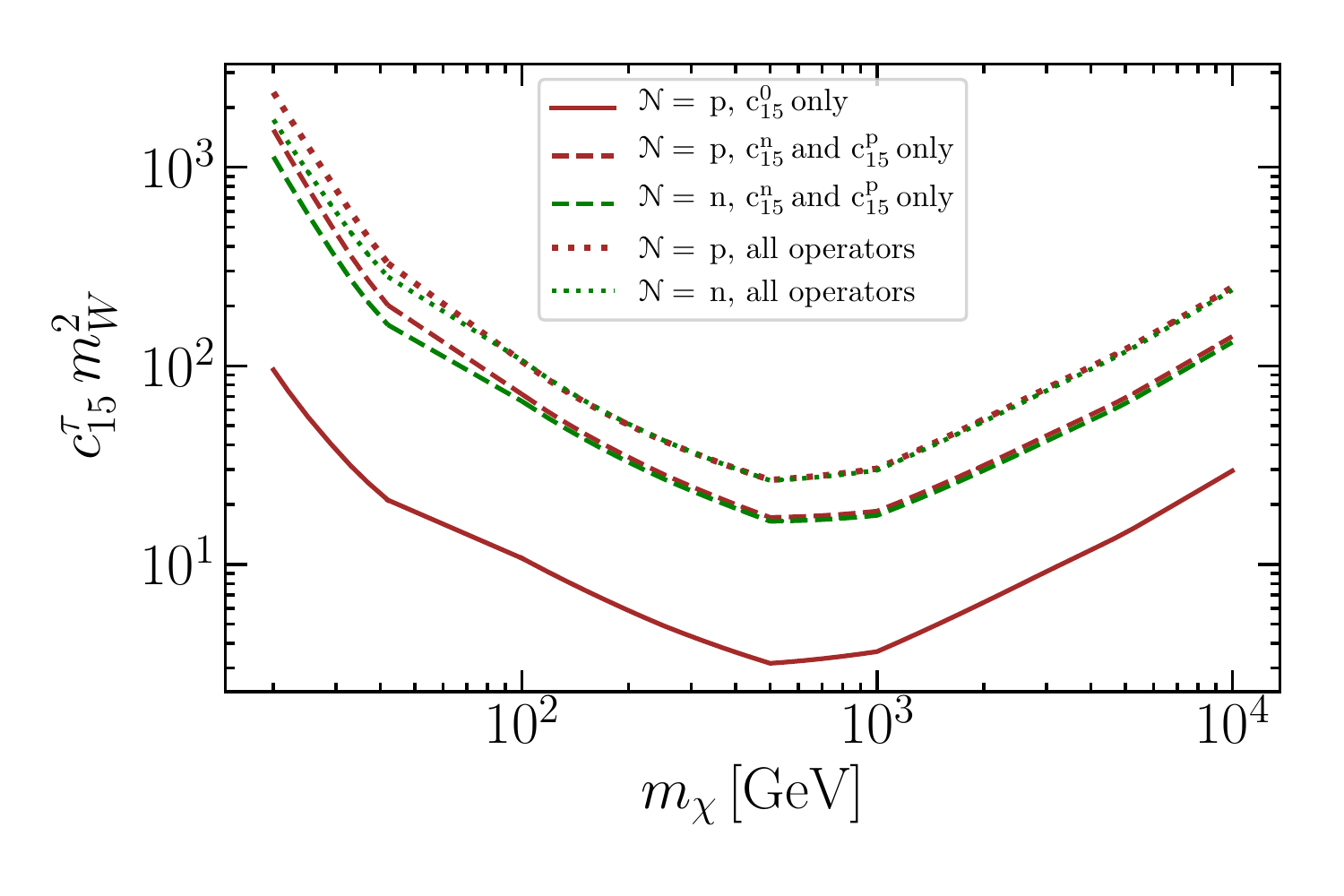}
	\end{center}
	\caption{\small Same as Fig. \ref{fig:UL-cnp-page1}, but for the operators $\widehat{\cal O}_i$, $i=8...15$.}
	\label{fig:UL-cnp-page2}
\end{figure}

We show in Figs.~\ref{fig:UL-cnp-page1} and \ref{fig:UL-cnp-page2} the upper limits on the coupling strengths in the neutron-proton basis. For each operator $\widehat{\cal O}_i$ the solid brown line shows the upper limit on the dark matter-nucleon coupling strength assuming that the only non-vanishing coupling strength is the isoscalar component $c_i^0$ (as commonly assumed in the literature). The dashed brown (green) line shows the upper limit on the dark matter-proton(neutron) coupling strength assuming that both $c_i^p$ and $c_i^n$ can be non-zero and can interfere, while all other coupling strengths vanish; and the dotted brown (green) line shows the upper limit on the dark matter-proton(neutron) coupling strength assuming that all coupling strengths can be non-zero.	Again, the impact of the interference can be very significant in some cases, and can relax the upper limits on the coupling strength by more than one order of magnitude. Let us stress that points above the dotted lines are necessarily ruled out, and therefore indicate conservative upper limits on the coupling strengths. Further, it is apparent from comparing Figs. \ref{fig:UL-c-page1}, \ref{fig:UL-c-page2} and Figs. \ref{fig:UL-cnp-page1}, \ref{fig:UL-cnp-page2}  that the effect of the interference is strongly basis dependent. This is also apparent from Fig.~\ref{fig:ellipse_bases}, and can be traced back to the orientation of the major axes of the ellipsoid ${\bf c}^T \mathbb{C} {\bf c} \leq C^{\rm u.l.}$ with respect to the axes of the isoscalar-isovector basis or the axes of the neutron-proton basis.

\begin{figure}[h!]
	\begin{center}
		\includegraphics[width=.45\textwidth]{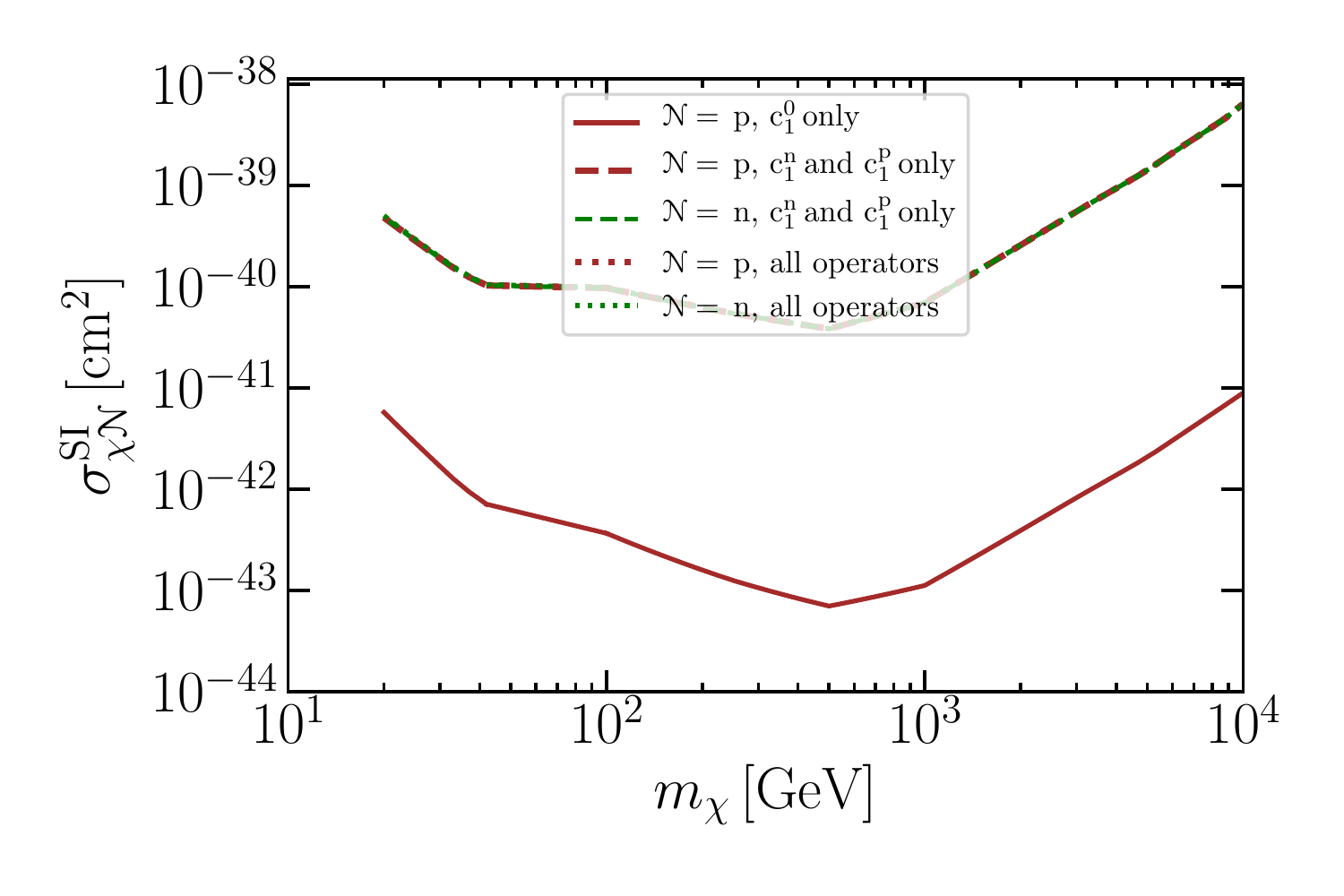}
		\includegraphics[width=.45\textwidth]{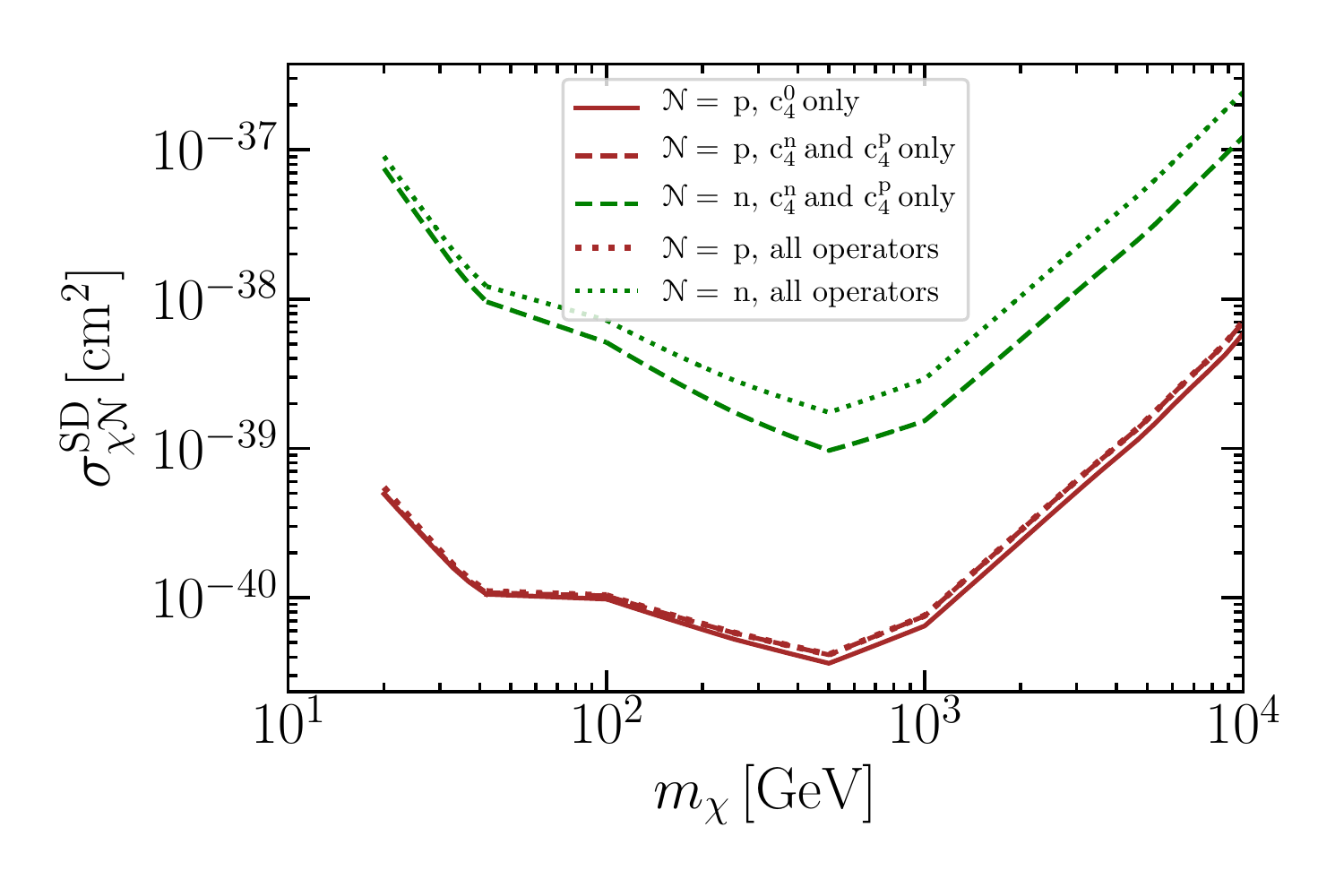}
	\end{center}
	\caption{\small Upper limits at the 90\% C.L. on the spin-independent (left panel) and spin-dependent (right panel) dark matter-nucleon scattering cross-section, for the annihilation final state $W^+W^-$ ($\tau^+\tau^-$ for $m_\chi<100$ GeV), assuming only one non-vanishing interaction (solid), the interference between the proton and neutron interactions (dashed) and the interference between all interactions of the effective theory (dotted).	
	}
	\label{fig:UL-sigma-page1}
\end{figure}

The IceCube collaboration presents the results of their search in terms of the SI and SD dark matter-proton cross-sections, assuming that the interaction is isoscalar, and assuming for the SD interaction that the dark matter has spin 1/2. In order to compare with published results, we show in Fig. \ref{fig:UL-sigma-page1} the effect of the interference among interactions for the SI (left panel) and the SD (right panel) dark matter-proton and dark matter-neutron interaction, translating the constraints on $c_1^{\mathscr N}$ and $c_4^{\mathscr N} $ from  Fig. \ref{fig:UL-cnp-page1}, into constraints on the cross-section, using 
\begin{align}
	\sigma^{\rm SI}_{\chi{\mathscr N}} &=\frac{\mu_{\chi{\mathscr N}}^2 (c_1^{\mathscr N})^2}{\pi}\;,\nonumber\\
	\sigma^{\rm SD}_{\chi{\mathscr N}} &=\frac{3}{16}\frac{\mu_{\chi{\mathscr N}}^2 (c_4^{\mathscr N})^2}{\pi}\;.
	\label{eq:x-section-nucleon}
\end{align}
Here, ${\mathscr N}=n,p$, and $\mu_{\chi\mathscr N}$ is the reduced mass of the dark matter-nucleon system. 

Strictly, values of the cross-section above the solid red line are ruled out {\it only} for models where the dark matter couples with equal strength to protons and neutrons. 
The upper limit applicable to {\it all} models where the dark matter interacts with the nucleon via the $\widehat{\cal O}_1$ (or $\widehat{\cal O}_4$) interaction, both isoscalar and isovector, corresponds to the dashed line,  and the upper limit applicable to {\it all} models corresponds to the dotted line, which are much weaker than the IceCube limit. Concretely, 
for the SI interaction, the limits on the dark matter-proton and the dark matter-neutron can be up to two orders of magnitude weaker than the ones derived under the assumption that the interaction is isoscalar. For the SD interaction, the interference barely affects the upper limits on the dark matter-proton cross-section, while it can relax the limits on the dark matter-neutron cross-section by up to two orders of magnitude.  This result could have implications to assess the viability of concrete models.

\afterpage{\FloatBarrier}

\section{Conclusions}
\label{sec:conclusions}

The non-relativistic scattering of dark matter particles off a nucleon doublet can be induced by various operators, involving all possible Galilean invariant combinations of the momentum transfer and relative velocity of the scattering, as well as the nucleon and dark matter spins. Further, the number of operators must be folded by the number of internal degrees of freedom of the nucleon doublet, namely the isospin, which can be described either in terms of a isoscalar and isovector component, or in terms of a neutron and proton component. The non-observation of exotic scattering signals allows to set upper limits on the coupling strength of the various interactions. However, due to the large number of possible interactions in the effective field theory (8 for spin-0 dark matter and 28 for spin-1/2), it is common to cast the limits on the scattering rate considering just one coupling strength, assuming that all others vanish. 

It is well known that operators in the effective theory with identical symmetry transformations under P, C and T in general interfere with one another, possibly destructively. Therefore, the limits on the coupling strength  that have been derived neglecting the interference among operators (and related quantities, such as the dark matter-proton and the dark matter-neutron cross-sections) can be too aggressive. Accordingly, the comparison of these limits with the predictions of a given model should be done with great care, as many models lead to more than one operator in the effective theory.

In this paper we have presented a method to derive a conservative upper limit on a coupling strength in the presence of interferences among operators. It exploits the fact that the scattering rate is a quadratic form in the coupling strengths of the effective theory. Therefore, the region of parameter space allowed by one experiment can be geometrically interpreted as the interior of an ellipsoid. We have derived an analytic formula to calculate the maximal value of one coordinate on the surface of the ellipsoid, which corresponds to the maximal value of the coupling strength permitted by that experiment. We have also discussed how to rederive the limits on the coupling strengths in a different basis. The basis transformation  can be applied for instance to determine conservative upper limits on the coupling strengths to neutrons and protons, when the dark matter scattering rate with a nucleus is calculated in terms of the isoscalar and isovector interactions of the dark matter with the nucleon, as commonly done in the literature. 

We have applied our method to derive conservative limits on the 14 isoscalar and isovector coupling strengths of the effective field theory of dark matter-nucleon interactions, restricting ourselves to dark matter spin up to 1/2. For each coupling strength, we have included the effect of the  interference between the isoscalar and isovector interactions of the same operator, or the interference between all operators. The effect of the interference relaxes the upper limits on the coupling strengths, in some cases by a factor $\sim 3$. We have also calculated conservative limits on the coupling strength of the 14 dark matter-proton and dark matter-neutron interactions. Again in this case the effect of the interference can be significant, and can relax the limits obtained under the common assumption that the interaction is isoscalar by up to one order of magnitude.  It is worth noting that, in general, the effect of the interference in deriving the limits differs from basis to basis.  
	
Finally, we have translated our results for the ${\cal O}_1$ and ${\cal O}_4$ interactions into conservative limits on the spin-independent and spin-dependent dark matter-nucleon interaction cross-sections, and we have compared with the published limits by the IceCube collaboration that assume that the interactions are isoscalar. For spin-independent interactions, the interference among operators can relax the upper limit on the dark matter-proton and the dark matter-neutron cross-sections by up to two orders of magnitude, compared to the upper limit derived under the common assumption that the interaction is isoscalar. For spin-dependent interactions, the interference among operators can relax the dark matter-neutron cross-section also by up to two orders of magnitude, whereas the limits for the dark matter-proton cross-section roughly coincide with the limit obtained by assuming isoscalar interactions.

The method can also be applied to direct detection experiments, and extended to obtain combined results from various experiments. The results will be presented in a forthcoming publication~\cite{inprep}.

\section*{Acknowledgements}
This work has been supported by the Collaborative Research Center SFB1258 and by the Deutsche Forschungsgemeinschaft (DFG, German Research Foundation) under Germany's Excellence Strategy - EXC-2094 - 390783311. We are grateful to Riccardo Catena, Gonzalo Herrera, Elisa Resconi, Gaurav Tomar and Martin Wolf for discussions and comments.

\appendix

\section{Complementarity of targets for capture inside the Sun}
\label{sec:complementarity}

For a given target nucleus, there are combinations of the isoscalar and isovector coupling strengths for which the recoil rate is suppressed. For this reason, it is of utmost importance to undertake a search for dark matter in direct detection experiments using multiple targets, so that no corner in the vast parameter space of the effective field theory remains unexplored. Interestingly, the Sun can be regarded as a multi-target dark matter detection experiment, and as we will argue below it is highly efficient in probing combinations of parameters that would only be very weakly constrained in a human-made experiment using a single element as target. 

\begin{figure}
	\begin{center}
		\includegraphics[width=.49\textwidth]{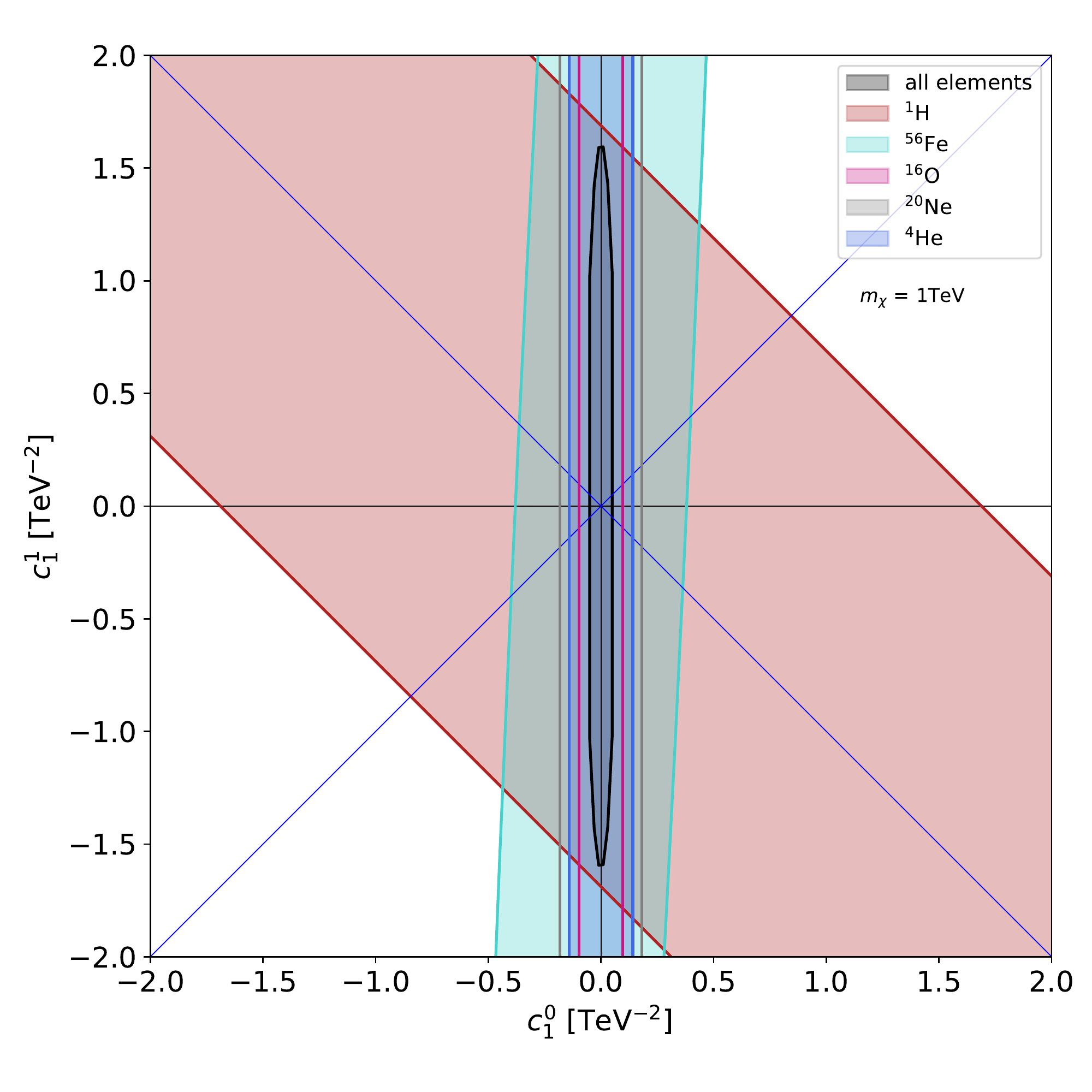}
		\includegraphics[width=.49\textwidth]{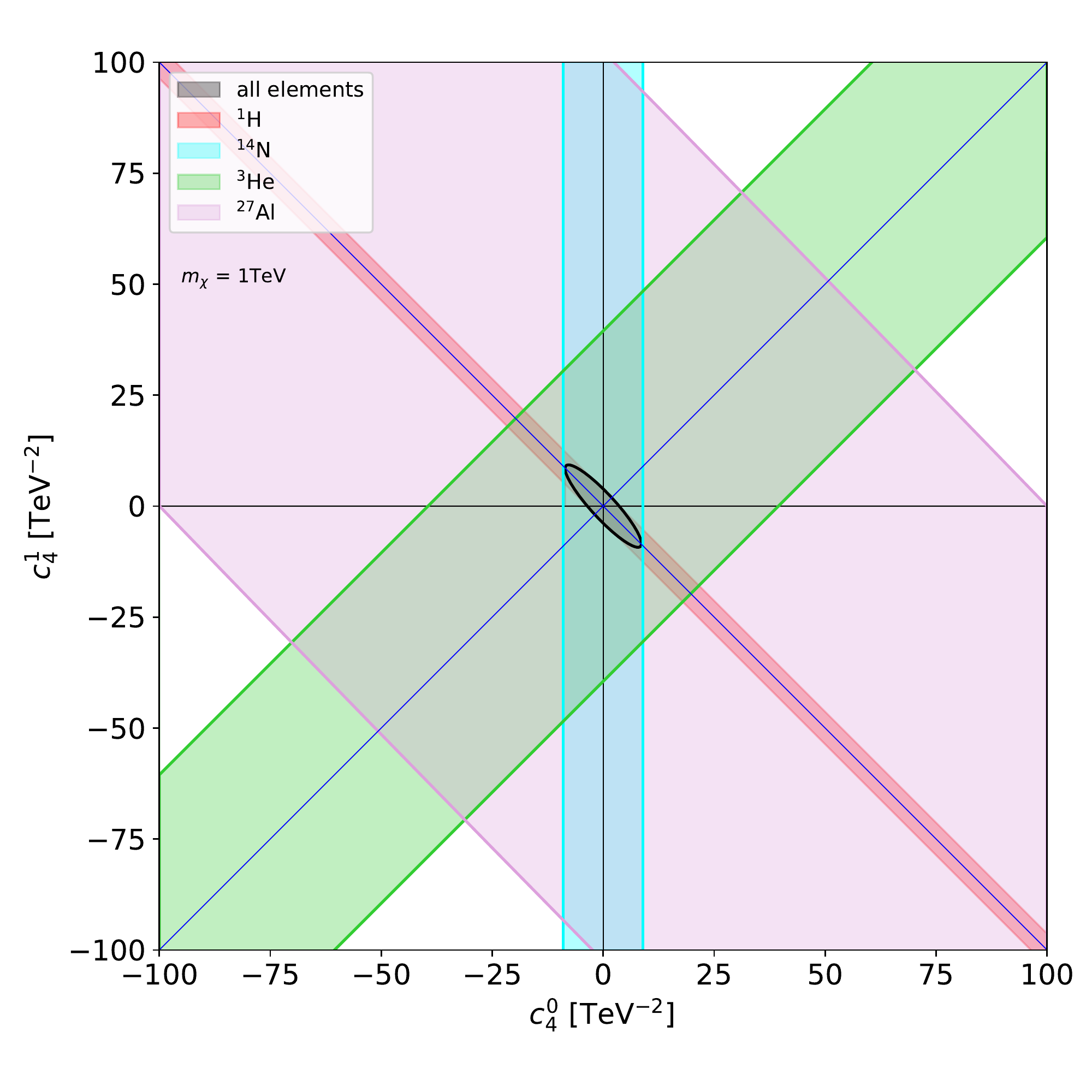}
	\end{center}
	\caption{\small  Regions of the isoscalar-isovector parameter space of the $\widehat{\cal O}_1$ and $\widehat{\cal O}_4$ operators (left and right panel, respectively) which are allowed by the current upper limit on the capture rate  from IceCube for $m_\chi=1$ TeV and the annihilation final state $W^+W^-$, assuming that the Sun contains only one element with mass density distribution given by the AGSS09ph solar model, and including all elements. }
	\label{fig:complementarity_SI_SD}
\end{figure}

This is illustrated in Fig.~\ref{fig:complementarity_SI_SD} which shows, for $m_\chi=1$ TeV, the regions of the isoscalar-isovector parameter space allowed by the current upper limit from IceCube on the capture rate, for the ${\cal O}_1$  and ${\cal O}_4$ operators (left and right panel, respectively),
assuming that the Sun contains only one element with mass density distribution given by the AGSS09ph solar model, and including all elements. Concretely, for the  ${\cal O}_1$ operator we show the region of the parameter space probed by $^1$H, $^{56}$Fe, $^{16}$O, $^{20}$Ne and $^4$He (other elements do not set any significant constraint on the region of the parameter space $|c_1^\tau|\leq 2\,{\rm TeV}^{-2}$). $^1$H probes only the dark matter-proton coupling, therefore, the corresponding ellipse is aligned with the $c_1^n=(c_1^0-c_1^1)/2$ axis. $^4$He,  $^{16}$O and $^{20}$Ne are nuclei with the same number of protons and neutrons, and therefore only probe the isoscalar interaction; the corresponding ellipses are then aligned with the $c_1^1$ axis. $^{56}$Fe contains 26 protons and 30 neutrons, and therefore probes mostly to isoscalar interaction, and to a lesser extent the isovector interaction. Likewise,  for  the  ${\cal O}_4$ operator we show the region of the parameter space probed by $^1$H, $^{14}$N, $^{3}$He and $^{27}$Al, which are the ones providing relevant constraints for $|c_4^\tau|\leq 100\,{\rm TeV}^{-2}$.  $^1$H and $^{27}$Al both contain one unpaired proton and therefore probe mostly the spin-dependent dark matter coupling to the proton. In contrast, $^{3}$He contains one unpaired neutron and correspondingly probes the spin-dependent dark matter coupling to the neutron. Lastly, $^{14}$N contains equal number of protons and neutrons and therefore probes the spin-dependent dark matter isoscalar coupling to the nucleon. Similar arguments can be applied to determine the dependence on the isoscalar or isovector couplings of the various interactions of the effective theory to the different elements inside the Sun.

\begin{figure}
	\begin{center}
		\includegraphics[width=\textwidth]{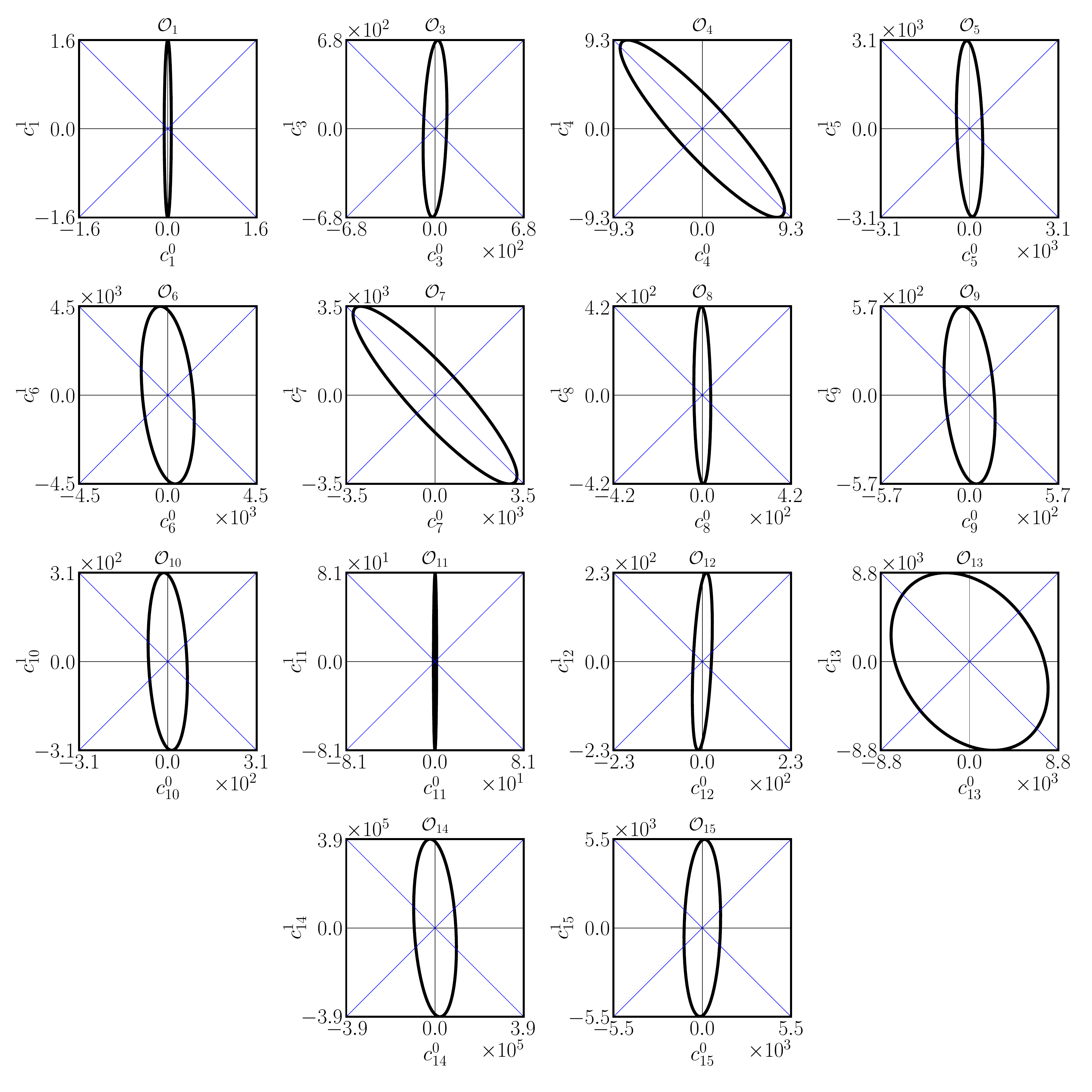}
	\end{center}
	\caption{\small	Values of the coupling strengths $c_i^0$ and $c_i^1$ (in units of TeV$^{-2}$)  for the operators $\widehat{\cal O}_i$, $i=1...14$, which are allowed by the current upper limit on the capture rate from IceCube assuming the annihilation final state $W^+W^-$ and $m_\chi=1$ TeV. }
	\label{fig:combined_ellipse_all_operator}
\end{figure}

As apparent from the plots, the presence of various elements in the Sun leads to an allowed region which is considerably smaller than the one allowed by each element individually. Further, while the orientation of the ellipses corresponding to each of the elements can be anticipated from the shell-model, the orientation of the combined ellipse depends on the dark matter mass and on the relative distribution of the elements inside the Sun, and can only be determined numerically. We show for completeness in Fig.~\ref{fig:combined_ellipse_all_operator} the orientation of the ellipse in the $c_i^0$-$c_i^1$ parameter space for the whole set of effective operators, again for $m_\chi=1$ TeV. Projecting the vertex and the co-vertex of the ellipse on the $c_i^0-c_i^1$ axes (or the rotated $c_i^n-c_i^p$ axes) one obtains the conservative limits on the strength of the different interactions in the isoscalar-isovector basis (or the neutron-proton basis).

\section{Upper limits on the coupling constants in the non-equilibrium regime}
\label{sec:no_equil}

In Section \ref{sec:method} we assumed that capture and annihilation are in equilibrium in the Sun. Here we extend the formalism for the case where this assumption does not hold, and which could be relevant for the theoretical interpretation of the neutrino flux from the Sun from annihilations as the constraints on the operators become more and more stringent, as well as for the analysis of the implications of the annihilation of captured dark matter particles in other objects. 

Using Eq.~(\ref{eq:relation_ann_cap_rate}) one can recast the annihilation rate today as a function of the capture rate:
\begin{align}
	\GammaA(C)= \frac{C}{2}\tanh^2(\kappa \sqrt{C})\;,
\end{align}
with $\kappa=t_\odot\sqrt{C_A}$ and $C = {\bf c}^T \mathbb{C} {\bf c}$. The non-observation of a neutrino flux in the direction of the Sun implies, for a given annihilation channel, $\GammaA(C)\leq \GammaA^{\rm u.l.}$. Following Section \ref{sec:method}, we can now determine the maximum value of the coupling constant $c_\alpha$ from the Lagrangian 
\begin{align}
	L=c_i-\lambda \Big[\GammaA(C)-\GammaA^{\rm u.l.}\Big]\;,
\end{align}
with $\lambda$ a Lagrange multiplier. The optimization conditions are
\begin{align}
	\frac{\partial L}{\partial c_\beta}\Big|_{{\bf c}={\bf c}^{\rm max}}\,&=\,\delta_{\alpha\beta}\,-\,2\lambda \frac{\partial \GammaA(C)}{\partial C}\Big|_{{\bf c}^{\rm max}}\mathbb{C}_{\beta\zeta} c^{\rm max}_\zeta \,=\,0\;,
	\label{eq:max_xi}\\
	\frac{\partial L}{\partial\lambda}\Big|_{{\bf c}={\bf c}^{\rm max}}\,&=
	\GammaA[C({\bf c}^{\rm max})]-\GammaA^{\rm u.l.}=0.
	\label{eq:max_lambda}
\end{align}

From Eq.~(\ref{eq:max_xi}) one obtains 
\begin{align}
	c_\beta^{\rm max}\,=\,\xi \left(\mathbb{C}^{-1}\right)_{\alpha\beta}\;,
	\label{eq:c-max-noequil}
\end{align}
with $\xi=(2\lambda \partial \GammaA(C)/\partial C|_{{\bf c}^{\rm max}})^{-1}$.
Substituting in Eq.(\ref{eq:max_lambda}) we find
\begin{align}
	\frac{1}{2}\xi^2\left(\mathbb{C}^{-1}\right)_{\alpha\alpha}\tanh^2\Big( \kappa\xi \sqrt{\left(\mathbb{C}^{-1}\right)_{\alpha\alpha}}\Big)=\GammaA^{\rm u.l.}.
	\label{eq:rel_C_Gammaul}
\end{align}
The root $\xi_0$ can be calculated numerically, given the matrix $\mathbb{C}$ and given the annihilation constant $\CA$ (related to the annihilation rate $\left<\sigma v\right>$ through Eq.~(\ref{eq:CA_approx})). Finally, the maximum value of the coupling constant $c_\alpha$ is given by
\begin{align}
	c_\alpha^{\rm max}\,=\,\xi_0 \left(\mathbb{C}^{-1}\right)_{\alpha\alpha}.
	\label{eq:c-max}
\end{align}

Alternatively, the maximum coupling constant $c_\alpha^{\rm max}$ can be obtained from the equation
	\begin{align}
		c_\alpha^{\rm max} \tanh\Big( \frac{\kappa}{\sqrt{(\mathbb{C}^{-1}})_{\alpha\alpha}}c_\alpha^{\rm max}\Big)=\sqrt{2(\mathbb{C}^{-1})_{\alpha\alpha}\GammaA^{\rm u.l.}}
		\label{eq:xi_0-II}
	\end{align}
which follows straightforwardly from Eq.~(\ref {eq:c-max}) and Eq.~(\ref{eq:rel_C_Gammaul}).

In the commonly assumed case that equilibration (such that $\tanh( t_0/\tau)\simeq 1$),
\begin{align}
	c_\alpha^{\rm max, eq}\,\simeq \sqrt{2\GammaA^{\rm u.l.} \left(\mathbb{C}^{-1}\right)_{\alpha\alpha}}.
\end{align}

With this definition, Eq.~(\ref {eq:xi_0-II}) can be recast as
\begin{align}
	\frac{c_\alpha^{\rm max}}{c_\alpha^{\rm max,eq}}\tanh\Big(\Xi\frac{c_\alpha^{\rm max}}{c_\alpha^{\rm max,eq}}\Big)=1\;,
	\label{eq:c_over_ceq}
\end{align}
where
\begin{align}
		\Xi=\kappa\sqrt{2\GammaA^{\rm u.l.} }=t_\odot\sqrt{2C_A\GammaA^{\rm u.l.}}=\frac{t_\odot}{\tau}\sqrt{\frac{2\GammaA^{\rm u.l.}}{C}}\;.
		\label{eq:Xi}	
	\end{align}

\begin{figure}
	\begin{center}
		\includegraphics[width=.49\textwidth]{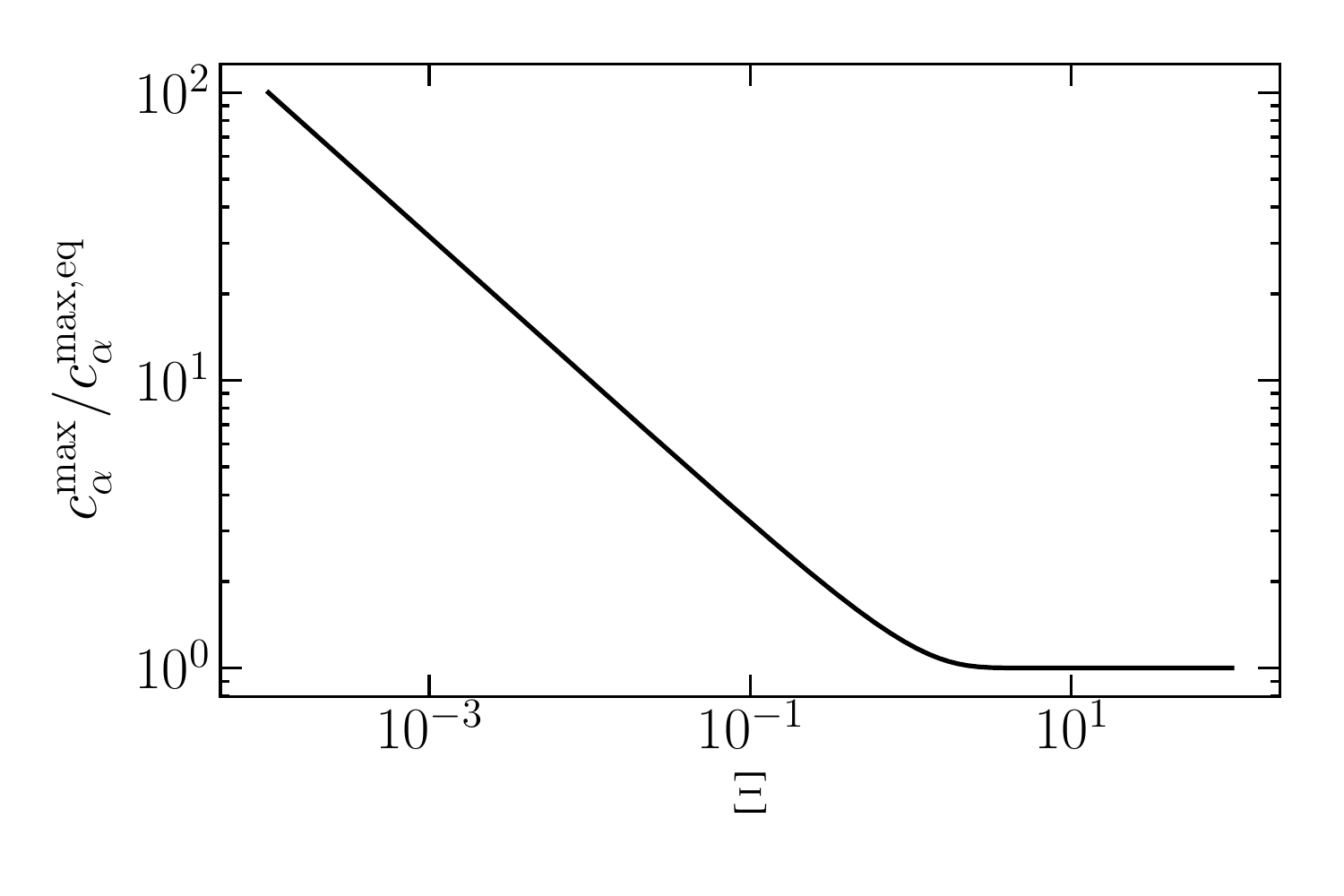}
		\includegraphics[width=.49\textwidth]{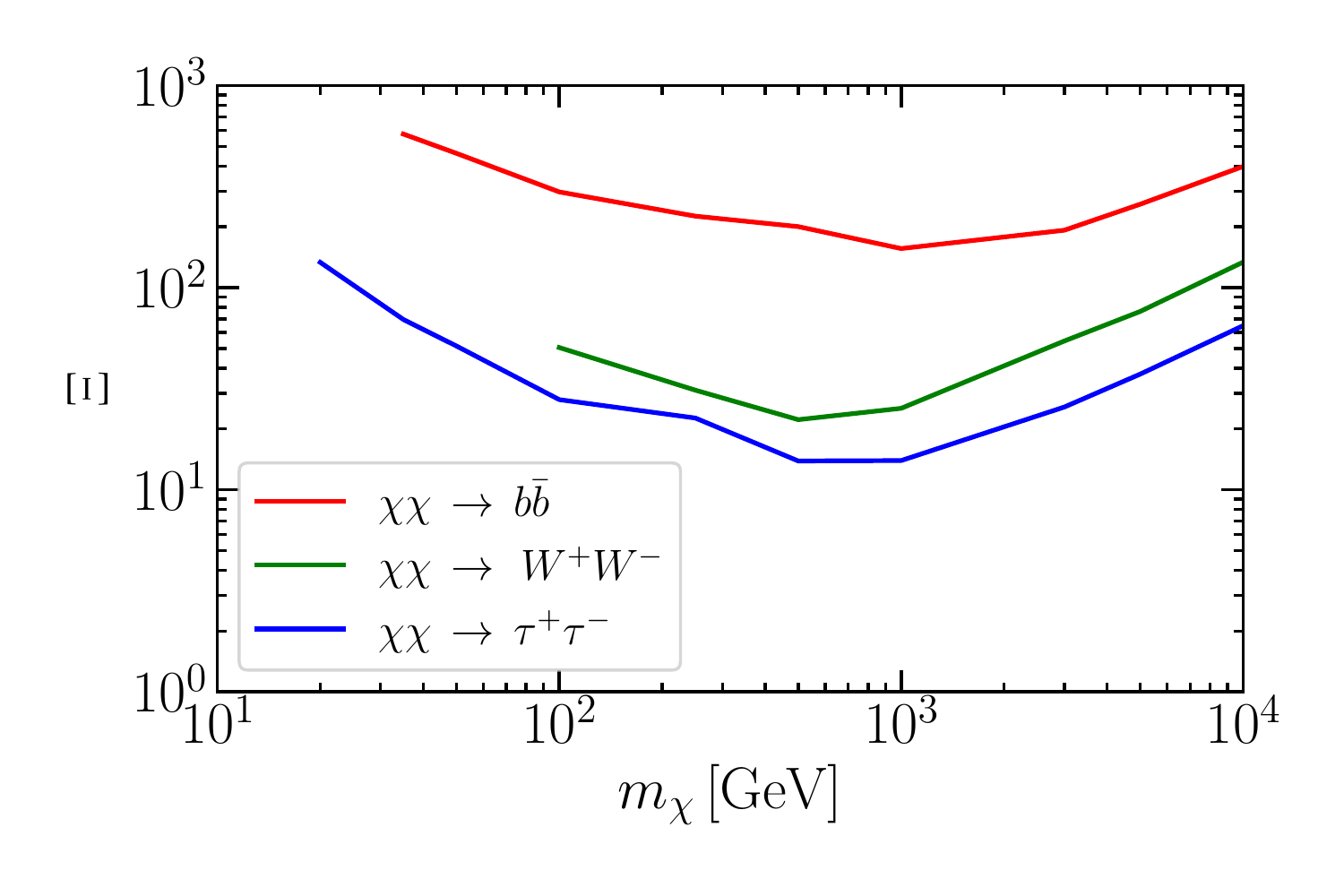}
	\end{center}
	\caption{\small {\it Left panel}: Change in the upper limit on any coupling strength of the effective theory of dark matter-nucleon interactions with respect to the corresponding upper limit derived under the assumption of equilibration between capture and annihilation, as a function of the parameter $\Xi$ defined in Eq.~(\ref{eq:Xi}). {\it Right panel}: Value of the parameter $\Xi$  as a function of the dark matter mass for the annihilation final states $b\bar b$ (red), $W^+W^-$ (green) and $\tau^+\tau^-$ (blue), assuming $(\sigma v)=3\times 10^{-26}\,{\rm cm}^3\,{\rm s}^{-1}$ and adopting the current upper limit from IceCube on the high energy neutrino flux from the Sun. For $(\sigma v)=3\times 10^{-26}\,{\rm cm}^3\,{\rm s}^{-1}$ and for the current limits from IceCube, one finds $\Xi\gg 1$ and therefore $c_\alpha^{\rm max}\simeq c_\alpha^{\rm max, eq}$. }
	\label{fig:Xi}
\end{figure}

In Fig.~\ref{fig:Xi}, left panel,  we show $c_\alpha^{\rm max}/c_\alpha^{\rm max,eq}$ as a function of $\Xi$ from the relation Eq.~(\ref{eq:c_over_ceq}). As expected, for $\Xi\gg 1$, equilibrium is reached and $c_\alpha^{\rm max}/c_\alpha^{\rm max,eq}\simeq 1$. For $\Xi \lesssim 1$, the upper limit on the coupling constant $c_\alpha^{\rm max}$ becomes weaker than the limit derived under the assumption of equilibration $c_\alpha^{\rm max,eq}\simeq 1$. This ``penalty factor" due to the lack of equilibration is independent on the considered operator and can be related to the annihilation cross-section and to the upper limit on the annihilation rate from the experiment through the second equality in Eq. ~(\ref{eq:Xi}). Using Eq.~(\ref{eq:CA_approx}) and assuming $(\sigma v)=3\times 10^{-26}\,{\rm cm}^3{\rm s}^{-1}$ one can derive the value of $\Xi$ as a function of the dark matter mass from the current upper limit on the annihilation rate reported by the IceCube experiment; the result is shown in Fig.~\ref{fig:Xi}, right panel, for the three exemplary annihilation channels used by IceCube. Under these assumptions $\Xi\gg 1$, and therefore the strict upper limits on the coupling constants, ${\rm max}\{c_\alpha\}$ are approximately equal to the ones derived from the hypothesis of equilibration between capture and annihilation. We note that $\Xi$ scales as $(\sigma v)^{1/2}$ and as $(\GammaA^{\rm u.l.})^{1/2}$.~\footnote{It is interesting to note that determining whether equilibration is attained or not is independent on the upper limit on $c_\alpha$ from direct detection experiments.} Therefore, for dark matter models where the annihilation cross-section inside the Sun is much smaller than the equilibrium value and/or if future neutrino telescopes improve the upper limit on $\GammaA$, the hypothesis of equilibration may not hold for some range of dark matter masses, and the inclusion of the ``penalty factor" in the derivation of the limits will become essential for assessing the viability of that scenario.

\FloatBarrier
\bibliographystyle{JHEP-mod}
\bibliography{references}

\end{document}